\documentclass[12pt]{article}
\usepackage{epsfig,amssymb,amsmath,psfrag,hyperref}
\usepackage[utf8]{inputenc}
\usepackage[english]{babel}
\usepackage{enumitem}
\usepackage{tikz}
\usepackage{color}
\usetikzlibrary{decorations.pathmorphing}
\usetikzlibrary{decorations.markings}
\usetikzlibrary{intersections}
\usetikzlibrary{calc}
\usetikzlibrary{positioning}
\usetikzlibrary{shapes}
\usetikzlibrary{fit}
\usetikzlibrary{backgrounds}
\hypersetup{
	linktocpage,
	colorlinks  = true, 
	urlcolor    = blue, 
	linkcolor   = blue, 
	citecolor   = red 
}

\def \be  {\begin{equation}}
\def \ee  {\end{equation}}
\def \ba  {\begin{eqnarray}}
\def \ea  {\end{eqnarray}}
\def \cO{\mathcal{O}}
\newcommand \widebar [1] {\overline{#1}}
\def\x{x}
\def\xb{\bar{x}}
\def\dbar#1{\widebar{D}_{#1}}
\def\ap{\alpha'}
\def\l{\lambda}
\def\fourtwo{\langle\cO_2\cO_2\cO_2\cO_2\rangle}
\def\L{\mathcal{L}}
\def\P{\mathcal{P}}
\def\d8{\Delta^{(8)}}
\def\ca{\mathcal{A}}

\begin{document}
\thispagestyle{empty}

\null\vskip-12pt \hfill  \\
\null\vskip-12pt \hfill   \\

\vskip 0.5truecm
\begin{center}
\vskip 0.2truecm {\Large\bf
{\Large Two-loop supergravity on AdS$_5\times$S$^5$ from CFT}
}\\
\vskip 1.25truecm
	{\bf J.~M.~Drummond$^1$ and H.~Paul$^2$
	}
	\vskip 0.4truecm
	{\it
		${}^{1}$ School of Physics and Astronomy and STAG Research Centre, \\
		University of Southampton,
		Highfield,  SO17 1BJ \vskip .2truecm
	}
	{\it
		${}^{2}$ Universit\'e Paris-Saclay, CNRS, CEA, Institut de Physique  Th\'eorique, \\
		91191, Gif-sur-Yvette, France \vskip .2truecm
	}
\end{center}

\vskip 1truecm 
\centerline{\bf Abstract}\normalsize
We describe a construction of the two-loop amplitude of four graviton supermultiplets in AdS$_5\times$S$^5$. We start from an ansatz for a preamplitude from which we generate the full amplitude under the action of a specific Casimir operator. The ansatz captures a recent ansatz of Huang and Yuan and we confirm their result through similar constraints. The form of the result suggests that all ambiguities are captured by the preamplitude which determines the result up to tree-level ambiguities only. We identify a class of four-dimensional `zigzag' integrals which are perfectly adapted to describing the leading logarithmic discontinuity to all orders. We also observe that a bonus crossing symmetry of the preamplitude follows from the transformation properties of the Casimir operator. Combined with the zigzag integrals this allows us to construct a crossing symmetric function with the correct leading logarithmic discontinuities in all channels.

From the two-loop result we extract an explicit expression for the two-loop correction to the anomalous dimensions of twist-four operators of generic spin which includes dependence on (alternating) nested harmonic sums up to weight three. We also revisit the prescription of the bulk-point limit of AdS amplitudes and show how it recovers the full flat-space amplitude, not just its discontinuity. With this extended notion of the bulk-point limit we reproduce the scale-dependent logarithmic threshold terms of type IIB string theory in flat-space.

\medskip
\noindent                                   
\newpage
\setcounter{page}{1}\setcounter{footnote}{0}
\tableofcontents
\newpage
\section{Introduction}\setcounter{equation}{0}
The AdS/CFT correspondence relates bulk AdS scattering amplitudes to correlation functions in a boundary conformal field theory. The nature of bulk dynamics can therefore be effectively explored by making use of tools from conformal field theory. The archetypal example of the correspondence is that of $\mathcal{N}=4$ super Yang-Mills theory as the boundary CFT, describing type IIB superstrings in the AdS$_5\times$S$^5$ bulk. Many recent works have explored this theory with much focus on tree-level four-point amplitudes in the bulk and their one-loop corrections \cite{Rastelli:2016nze,Rastelli:2017udc,Alday:2017xua,Aprile:2017bgs}. Such bulk amplitudes correspond to a large $N$ expansion of the correlation function of four half-BPS operators in the conformal field theory.

Here we would like to explore the possibility of extending the analysis to higher loop order, focussing on two loops as a first example. This may seem a formidable problem, particularly from the bulk perspective, where already tree-level and one-loop amplitudes pose significant computational problems \cite{Freedman:1998tz,DHoker:1999kzh,DHoker:1999mqo,Arutyunov:2000py,Uruchurtu:2008kp,Uruchurtu:2011wh,Arutyunov:2017dti,Arutyunov:2018tvn,Yuan:2018qva,Carmi:2019ocp}. However, the end results for tree-level and one-loop amplitudes, obtained mostly from boundary CFT considerations, actually display surprising simplicity. In part, this is due to a hidden ten-dimensional conformal symmetry of the tree-level amplitudes which in turn leads to a significant simplification of the leading logarithmic discontinuity at all loop orders.

We will focus on the four-point scattering of graviton supermultiplets, or equivalently the four-point correlation function of four stress-energy multiplets in $\mathcal{N}=4$ super Yang-Mills theory. In \cite{Aprile:2019rep} it was observed that the one-loop result, first computed in \cite{Aprile:2017bgs}, can actually be simplified significantly with the help of an eighth-order conformal Casimir operator. This operator was known to simplify the form of the leading logarithmic discontinuity to all loop orders \cite{Aprile:2018efk}. It is related to the surprisingly simple form of the tree-level contribution to the anomalous dimensions of the double-trace operators exchanged in the OPE with two external half-BPS operators \cite{Aprile:2017xsp,Aprile:2018efk}. The simple form of the anomalous spectrum is in turn related to the ten-dimensional conformal symmetry of the tree-level amplitudes \cite{Caron-Huot:2018kta}. With the use of the eight-order Casimir the one-loop amplitude can be replaced by a much simpler preamplitude, whose analytic structure is so tightly constrained that in fact one does not even require the form of the leading discontinuity to essentially fix it uniquely~\cite{Aprile:2019rep}.

Here, in common with the recent work of Huang and Yuan \cite{Huang:2021xws}, we explore the extension of the use of this operator to higher loops. In particular we write the amplitude in terms of a preamplitude from which we obtain the full amplitude by two applications of the eighth-order Casimir. We observe that the square of the Casimir actually exhibits a simple transformation under crossing symmetry which allows us to restrict the preamplitude to be fully crossing symmetric. We describe an anzatz for the preamplitude in terms of polylogarithmic functions of two variables, including singularities which appear in the one-loop string corrections to the supergravity amplitude \cite{Drummond:2019hel}. Then we impose various OPE predictions arising from known tree-level and one-loop data. Under the further assumption that the leading discontinuity is manifested by the preamplitude, and imposing that the amplitude correctly reproduces the two-loop flat-space supergravity amplitude upon taking the bulk-point limit, we are led to confirm the result of \cite{Huang:2021xws}. In particular, we fix some ambiguities left undetermined in their result and, as a consequence of the interplay of crossing symmetry with the canonical form of the leading discontinuity (as well as the conjectured absence of weight 4 functions containing the letter $x-\xb$), we provide a justification for the vanishing of their last free parameter. We also explore certain wider ans\"atze and analyse the corresponding freedoms they introduce into the final form of the two-loop amplitude.

The result we obtain is very suggestive that in fact all ambiguities should also be included in the form of the preamplitude ansatz, that is to say that any finite contributions from tree-level and one-loop counterterms should also respect the form of the ansatz given in terms of the preamplitude, as also happens at one loop. This possibility restricts the final amplitude to just a handful of ambiguities, which are exactly the expected tree-level ambiguities consistent with the presence of genus-two contributions from the $\partial^{4}\mathcal{R}^4$, $\partial^{6}\mathcal{R}^4$, $\partial^{8}\mathcal{R}^4$ and $\partial^{10}\mathcal{R}^4$ corrections. Notably, the $\mathcal{R}^4$ term does not appear as an independent ambiguity, in analogy with its vanishing contribution at genus two.

Furthermore, from our result we extract the two-loop correction to the anomalous dimensions of the twist-four double-trace operators $\cO_{2,\ell}\sim\cO_2\partial^{\ell}\cO_2$ for generic spin $\ell$. In the supergravity limit (and neglecting $1/\l$ corrections), their dimensions admit a large $N$ expansion of the form
\begin{align}
	\Delta_{2,\ell} = 4+\ell+2\big(a\gamma^{(1)}+a^2\gamma^{(2)} + a^3\gamma^{(3)}+O(a^4)\big),
\end{align}
where we introduced $a\equiv1/(N^2-1)$. To order $a^3$, the anomalous dimensions read
\begin{align}
\begin{split}\label{eq:gammas_summary}
	\gamma^{(1)} &= -\frac{48}{(\ell+1)(\ell+6)}\,,\\[3pt]
	\gamma^{(2)} &= \frac{1344 (\ell-7) (\ell+14)}{(\ell-1) (\ell+1)^2 (\ell+6)^2 (\ell+8)}-\frac{2304 (2 \ell+7)}{(\ell+1)^3 (\ell+6)^3} -\frac{1080}{7}\,\delta _{\ell,0} \,,\\[3pt]
	\gamma^{(3)} &= c_3\,\big(S_{-3}-S_{3}-2S_{1,-2}+3\zeta_3\big)+c_2\,S_{-2}+c_1\,S_1+c_0+c_{0}^{(a)}+\alpha\,\tilde{\gamma}^{(2,3)}_{2,\ell},
\end{split}
\end{align}
where the tree-level and one-loop contributions $\gamma^{(1)}$ and $\gamma^{(2)}$ have been derived in \cite{Hoffmann:2000dx} and \cite{Aprile:2017bgs}, respectively, while the two-loop contribution $\gamma^{(3)}$ constitutes a new result.\footnote{
	Note that the formula for $\gamma^{(3)}$ quoted above is strictly speaking only valid for spins $\ell\geq6$. This is because we are able to fix the two-loop amplitude only up to certain tree-level ambiguities which contribute up to spin $\ell=4$ in the conformal block expansion.
	}
The coefficients $c_i$ are rational functions of spin and $S_{\vec{a}}\equiv S_{\vec{a}}(\ell+3)$ are nested harmonic sums, whose precise definitions as well as an extended discussion of the properties of $\gamma^{(3)}$ are given in Section~\ref{sec:gamma3}.\\

The rest of the paper is organised as follows: the remainder of this section is devoted to fixing our notation for the $\fourtwo$ correlator and introducing its OPE decomposition. 

In Section~\ref{sec:leading-logs}, we focus on the leading logarithmic divergence of supergravity correlators to any loop-order. We notice that its transcendental structure is described by a certain family of four-dimensional loop-integrals, the so-called zigzag integrals, which are a particular class of the generalised ladders discussed in \cite{Drummond:2012bg}.
This observation is powerful as it determines a part of the supergravity correlator at any loop-order, greatly reducing the number of transcendental functions with undetermined coefficients.

Next, in Section~\ref{sec:bulk_point} we revisit the bulk-point limit of AdS amplitudes and we show how known tree-level and one-loop AdS$_5\times$S$^5$ amplitudes match their flat-space counterparts. Moreover, the extra $x-\xb$ singularity of the one-loop string corrections turns out to be a crucial ingredient which reproduces the logarithmic threshold terms present in the perturbative expansion of the type IIB string amplitude in flat-space. Extended to two-loop order, the knowledge of the 10-dimensional two-loop supergravity amplitude gives a non-trivial constraint for the corresponding AdS correlator.

The main result of our work is described in Section~\ref{sec:two_loops}, where we spell out the details on the minimal ansatz used as a starting point of our bootstrap approach to the two-loop supergravity contribution. In particular, we discuss in great detail the basis of transcendental functions and compare our result to the one recently obtained in~\cite{Huang:2021xws}.

In Section~\ref{sec:gamma3}, we make use of the obtained two-loop correlator to extract from it new unprotected CFT data, recall also the previous comments around equation~\eqref{eq:gammas_summary}. The result for the two-loop anomalous dimension extends the previously known terms up to order $a^3$.

Lastly, in Section~\ref{sec:wider_ansatz} we consider possible natural extensions of the minimal ansatz and analyse the corresponding additional free parameters in the final result. We then conclude with a number of open questions.

\subsection{Setup: the $\fourtwo$ correlator}
We consider the four-point correlation function of the stress-tensor superprimary $\cO_2$, which is a half-BPS single-trace operator of conformal dimension $\Delta=2$. It is given in terms of the fundamental scalar fields $\Phi_i$ ($i=1,\ldots,6$) of the $\mathcal{N}=4$ supermultiplet by 
\begin{align}
	\cO_2(x,y)= y^i y^j ~\text{Tr}\big(\Phi_i(x)\Phi_j(x)\big),
\end{align}
with $y^i$ being an auxiliary $so(6)$ null vector obeying $y\cdot y=0$, such that $\cO_2$ transforms in the traceless symmetric representation $[0,2,0]$ of the R-symmetry group $su(4)$. In the context of the AdS/CFT correspondence, this operator is dual to the scalar in the graviton multiplet of type IIB supergravity on an AdS$_5\times$S$^5$ background, whereas its higher charge cousins $\cO_p$ with $p\geq3$ are dual to Kaluza-Klein modes arising from the compact S$^5$ factor.\footnote{
	To be more precise, the exact form of `single-particle operators' $\cO_p$ which are dual to single-particle states in AdS supergravity in general contains admixtures of multi-trace operators, whose coefficients are $1/N$ suppressed with respect to the usual single-trace term~\cite{Aprile:2018efk}. For an extensive study of the properties of these single-particle operators and their free-theory correlators, see~\cite{Aprile:2020uxk}.
	}

We are interested in the four-point correlator
\begin{align}
	\langle\cO_2(x_1,y_1)\cO_2(x_2,y_2)\cO_2(x_3,y_3)\cO_2(x_4,y_4)\rangle=g_{12}^2g_{34}^2\,\mathcal{G}(u,v;\sigma,\tau),
\end{align}
where $g_{ij}=y_{ij}^2/x_{ij}^2$ (with $y_{ij}^2=y_i\cdot y_i$) are propagator factors which account for the conformal weight and the scaling weights $y_i$ of the correlator, such that $\mathcal{G}$ depends only on the conformal and $su(4)$ R-symmetry cross-ratios. These are defined by
\begin{equation}\label{eq:crossratios}
\begin{aligned}
	u= x \xb &= \frac{x_{12}^2x_{34}^2}{x_{13}^2 x_{24}^2},\qquad &&~v=(1-x)(1-\xb)=  \frac{x_{14}^2x_{23}^2}{x_{13}^2 x_{24}^2}, \\
	\frac{1}{\sigma}=y \bar y &= \frac{y_{12}^2 y_{34}^2}{y_{13}^2 y_{24}^2},\qquad &&\frac{\tau}{\sigma}=(1-y)(1-\bar y)=\frac{y_{14}^2 y_{23}^2}{y_{13}^2 y_{24}^2}.
\end{aligned}
\end{equation}
As a consequence of superconformal symmetry the function $\mathcal{G}$ is furthermore constrained to take the form~\cite{Eden:2000bk,Nirschl:2004pa}
\begin{align}\label{eq:partial_non-ren}
	\mathcal{G}(u,v;\sigma,\tau) = \mathcal{G}_{\text{free}}(u,v;\sigma,\tau)+\mathcal{I}~\mathcal{H}(u,v).
\end{align}
For convenience, we normalise the correlator by a factor of $(N^2-1)^2$ such that the free theory contribution is given by
\begin{align}\label{eq:free_2222}
	\mathcal{G}_{\text{free}}(u,v;\sigma,\tau) = 4\Big(1+u^2\sigma^2+\frac{u^2\tau^2}{v^2}\Big)+16a\Big(u\sigma+\frac{u\tau}{v}+\frac{u^2\sigma\tau}{v}\Big).
\end{align}
We call the second contribution to~\eqref{eq:partial_non-ren} the interacting or dynamical part, as it is the only piece of the correlator which depends on the gauge coupling $g_{\text{YM}}$ and hence contains all of the non-trivial dynamics of the theory. It is of the factorised form shown above with the factor $\mathcal{I}$ being fixed by the superconformal Ward identities,
\begin{align}\label{eq:intriligator_factor}
	\mathcal{I} = \frac{(x-y)(x-\bar{y})(\xb-y)(\xb-\bar{y})}{(y\bar{y})^2}\,.
\end{align}
As a feature of the $\fourtwo$ correlator, and more generally for next-to-next-to-extremal correlators, the function $\mathcal{H}$ is then independent of the $su(4)$ variables. Furthermore, it obeys the crossing symmetries
\begin{align}\label{eq:crossing_H}
	\mathcal{H}(u,v) = \frac{1}{v^2}\mathcal{H}(u/v,1/v) = \frac{u^2}{v^2}\mathcal{H}(v,u),
\end{align}
placing strong constraints on its functional form which we will exploit in our bootstrap approach.

We will study the function $\mathcal{H}$ in the supergravity regime (reached by first taking $N$ large with the 't Hooft coupling $\lambda=g_{\text{YM}}^2N$ being held fixed and then taking $\lambda$ to infinity), where it describes the scattering of four supergravitons in type IIB superstring theory on AdS$_5\times$S$^5$. Ignoring the $\l$-dependence for a moment and focussing only on the supergravity contributions, $\mathcal{H}$ admits the large $N$ expansion
\begin{align}\label{eq:large_N_expansion}
	\mathcal{H}(u,v) = a\,\mathcal{H}^{(1)}(u,v) + a^2\,\mathcal{H}^{(2)}(u,v) + a^3\,\mathcal{H}^{(3)}(u,v) + O(a^4)\,,
\end{align}
where for convenience we choose to expand in small $a$, recall the definition $a=1/(N^2-1)$.\footnote{
	In terms of the central charge $c$, one has $a=\frac{1}{4c}$.}
The first term in the expansion is given by the well known tree-level supergravity correlator $\mathcal{H}^{(1)}$~\cite{Arutyunov:2000py,Dolan:2001tt}, while the higher-order terms correspond to loop amplitudes in the bulk: the order $a^2$ term $\mathcal{H}^{(2)}$ has been computed in~\cite{Aprile:2017bgs}\footnote{See also references~\cite{Alday:2017xua} and~\cite{Alday:2017vkk}.} and is dual to the one-loop AdS amplitude. Next comes the two-loop correction $\mathcal{H}^{(3)}$ which is the main object of interest here, and has recently also been considered in~\cite{Huang:2021xws}.

Let us now comment on the structure of $\l$-dependent terms, which we have omitted in~\eqref{eq:large_N_expansion}. These terms arise due to higher-derivative corrections of the string theory effective action on AdS$_5\times$S$^5$, and in the flat-space limit they are related to analogous terms in the low-energy expansion of the ten-dimensional four-point scattering amplitude of massless string states in type IIB string theory. To this end, we will denote a contribution at order $a^n\l^{-\frac{m}{2}}$ by $\mathcal{H}^{(n,m)}$. To be consistent with the expansion~\eqref{eq:large_N_expansion}, we have $\mathcal{H}^{(n)}\equiv\mathcal{H}^{(n,0)}$.

At tree-level, such terms are given by contact Witten diagrams whose interaction vertices are higher-derivative corrections to supergravity at genus zero. These vertices are of the form $\partial^{2k}\mathcal{R}^4$ and give rise to the tree-level terms $\mathcal{H}^{(1,k+3)}$.\footnote{
	These string corrections have been addressed in~\cite{Alday:2018pdi,Binder:2019jwn,Drummond:2019odu,Drummond:2020dwr,Aprile:2020mus}.} 
Note that in general these terms receive higher-genus corrections which contribute to higher orders in $a$, while still being `tree-level' in the bulk.\footnote{
	The existence of such contributions ultimately follows from S-duality of type IIB string theory, see references~\cite{Green:1999pv,Chester:2019jas,Chester:2020vyz} for a study of modular invariance in this context and also the recent work~\cite{Collier:2022emf} for a new approach to $SL(2,\mathbb{Z})$ invariance in the context of $\mathcal{N}=4$ SYM.}

For instance, the $\mathcal{R}^4$ vertex receives a genus-one correction $\mathcal{R}^4\vert_{\text{genus-1}}$ at order $a^2\lambda^{\frac{1}{2}}$. This gives rise to a term $\mathcal{H}^{(2,-1)}$, which happens to be super-leading with respect to the one-loop supergravity correction $\mathcal{H}^{(2)}$. As argued for in~\cite{Aharony:2016dwx}, this super-leading term can be thought of as a finite counter-term from string theory which regularises the one-loop supergravity divergence in AdS$_5\times$S$^5$. The subsequent terms are then suppressed by powers of $1/\l$, corresponding to either genus-one corrections to the higher-derivative tree-level terms $\mathcal{H}^{(1,k+3)}$ or genuine one-loop contributions induced by their very presence at order $a$, which have been addressed in~\cite{Alday:2018kkw,Drummond:2019hel,Drummond:2020uni}.

Similarly, at order $a^3$, there exist genus-two contributions from certain tree-level contact terms which precede the two-loop supergravity correlator $\mathcal{H}^{(3)}$. These are the genus-two corrections to the $\partial^4\mathcal{R}^4$ and $\partial^6\mathcal{R}^4$ vertices, corresponding to the terms $\mathcal{H}^{(3,-3)}$ and $\mathcal{H}^{(3,-2)}$ in our notation. Next, at order $a^3\l^{\frac{1}{2}}$, a new effect takes place: there are two distinct contributions to $\mathcal{H}^{(3,-1)}$, namely the genus-two correction to the tree-level $\partial^8\mathcal{R}^4$ term and a one-loop contribution involving a supergravity and a $\mathcal{R}^4\vert_{\text{genus-1}}$ vertex. Finally, the genus-two correction to the $\partial^{10}\mathcal{R}^4$ term is at the same order as two-loop supergravity, and thus will also contribute to $\mathcal{H}^{(3)}$.

Lastly, let us remark that the coefficients of some of these super-leading terms have been computed up to genus three using supersymmetric localisation techniques, see~\cite{Chester:2019pvm,Chester:2020dja} for more details.

\subsection{OPE decomposition and double-trace spectrum}
Next, we introduce the OPE decomposition of the $\fourtwo$ correlator. In particular, we will be interested into its unprotected (or long) part, which receives contributions from both the free theory and the interacting part and admits a decomposition into superconformal blocks
\begin{align}\label{eq:block_deco}
\fourtwo_{\text{long}} = g_{12}^2g_{34}^2~\mathcal{I}~\sum_{t,\ell}A_{t,\ell}\,G_{t,\ell}(\x,\xb).
\end{align}
The sum above runs over all long superconformal primaries with half the twist $t\equiv\frac{\Delta-\ell}{2}$ and even spin $\ell$ which are present in the $su(4)$ singlet representation in the OPE of $\cO_2\times\cO_2$. The $A_{t,\ell}$ are the squared OPE coefficients and the functions $G_{t,\ell}(x,\xb)$ are related to the usual four-dimensional conformal blocks by a shift in their dimensions, giving~\cite{Dolan:2000ut,Dolan:2003hv}
\begin{align}\label{eq:conformal_block}
	G_{t,\ell}(\x,\xb) = (-1)^\ell (\x\xb)^t~\frac{\x^{\ell+1}F_{t+\ell+2}(\x)F_{t+1}(\xb)-\xb^{\ell+1}F_{t+\ell+2}(\xb)F_{t+1}(\x)}{\x-\xb},
\end{align}
with $F_{\rho}(\x)={}_2F_1\left(\rho,\rho,2\rho;\x\right)$ being the standard hypergeometric function.

To leading order in the expansion around the supergravity limit introduced above, the only unprotected operators exchanged in the OPE of $\cO_2\times\cO_2$ are singlet channel double-trace operators of even spin $\ell$. Such operators are constructed from products of two half-BPS operators, and we denote their classical dimensions by $\Delta^{(0)}=2t+\ell$. Note that generically there are many such operators with the same quantum numbers, leading to operator mixing: for a given half-twist $t$, there are $t-1$ such operators which we distinguish by their degeneracy label $i=1,\ldots,t-1$. Schematically, the set of degenerate double-trace operators is given by
\begin{align}\label{eq:double_trace_operators}
	\cO_{t,\ell,i} = \big\{\cO_2\square^{t-2}\partial^\ell\cO_2,~\cO_3\square^{t-3}\partial^\ell\cO_3,~\ldots~,~\cO_t\square^0\partial^\ell\cO_t\big\}.
\end{align}In the supergravity limit, their dimensions $\Delta_{t,\ell,i}$ and OPE coefficients $A_{t,\ell,i}\equiv\langle\cO_2\cO_2\cO_{t,\ell,i}\rangle^2$ admit the expansion
\begin{align}\label{eq:expansion_dimensions}
	\Delta_{t,\ell,i} &= \Delta^{(0)}+2\big(a\gamma^{(1)}_i+a^2\gamma^{(2)}_i+a^3\gamma^{(3)}_i+O(a^4)\big)\,,\\\label{eq:expansion_OPE_coeffs}
	A_{t,\ell,i}    &= A^{(0)}_{t,\ell,i} + aA^{(1)}_{t,\ell,i} + a^2A^{(2)}_{t,\ell,i} + a^3A^{(3)}_{t,\ell,i} + O(a^4)\,,
\end{align}
where we have again suppressed the $\l$-dependence at it is understood that the anomalous dimensions $\gamma^{(n)}_i$ themselves depend on $t$ and spin $\ell$.

In the singlet channel and up to order $a$, the mixing of the operators $\cO_{t,\ell,i}$ has been fully resolved~\cite{Aprile:2017bgs,Aprile:2017xsp}, leading to a compact expression for their tree-level anomalous dimensions\footnote{For a generalisation to double-trace operators in any $su(4)$ channel, see~\cite{Aprile:2018efk}.}
\begin{align}\label{eq:gamma1}
	\gamma^{(1)}_{i} = -\frac{2(t-1)_4(t+\ell)_4}{(\ell+2 i-1)_6}\,,
\end{align}
and leading-order OPE coefficients
\begin{align}\label{eq:A0}
	A^{(0)}_{t,\ell,i} = \frac{8(t+\ell+1)!^2 t!^2 (\ell+1)(2t+\ell+2)}{(2t)!(2t+2\ell+2)!}~R_{t,\ell,i}\,b_{t,i}\,,
\end{align}
where
\begin{align}
\begin{split}
	R_{t,\ell,i} &= \frac{2^{1-t}(2\ell+3+4i) [(\ell+i+1)_{t-i-1}]^{{\rm sign}(t-i-1)} (t+\ell+4)_{i-1}}{(\tfrac{5}{2}+\ell+i)_{t-1}}\,,\\
	b_{t,i} &= \frac{2^{(1 - t)} (2 + 2 i)! (t-2)! (2t - 2 i + 2)!}{3 (i-1)! (i+1)! (t+2)! (t-i-1)! (t-i+1)!}\,.
\end{split}
\end{align}
Note that the first factor in equation~\eqref{eq:A0} can be thought of as an averaged OPE coefficient $\langle A^{(0)}_{t,\ell} \rangle$ over the double-trace degeneracies $i$, since $\sum_{i=1}^{t-1}R_{t,\ell,i}\,b_{t,i}=1$.

To conclude this section, let us indicate how the double-trace OPE data introduced above is encoded within the supergravity correlators across different orders in $a$. By plugging in the expansions~\eqref{eq:expansion_dimensions} and~\eqref{eq:expansion_OPE_coeffs} into the superconformal block decomposition~\eqref{eq:block_deco}, one finds
\begin{align}\label{eq:logu_stratification}
\begin{split}
	\mathcal{H}^{(1)}&=~~~\log^1(u)\,\big[A^{(0)}\gamma^{(1)}\big]\,G_{t,\ell}(x,\xb)\\
	&~~~+\log^0(u)\,\big[A^{(1)}+2A^{(0)}\gamma^{(1)}\partial_{\Delta}\big]\,G_{t,\ell}(x,\xb)\,,\\
	\mathcal{H}^{(2)}&=~~~\log^2(u)\,\big[\tfrac{1}{2}A^{(0)}(\gamma^{(1)})^2\big]\,G_{t,\ell}(x,\xb)\\
	&~~~+\log^1(u)\,\big[(A^{(1)}\gamma^{(1)}+A^{(0)}\gamma^{(2)})+2A^{(0)}(\gamma^{(1)})^2\partial_{\Delta}\big]\,G_{t,\ell}(x,\xb)\\
	&~~~+\log^0(u)\,\big[A^{(2)}+2(A^{(1)}\gamma^{(1)}+A^{(0)}\gamma^{(2)})\,\partial_{\Delta}+2A^{(0)}(\gamma^{(1)})^2\partial_{\Delta}^2\big]\,G_{t,\ell}(x,\xb)\,,\\
	\mathcal{H}^{(3)}&=~~~\log^3(u)\,\big[\tfrac{1}{6}A^{(0)}(\gamma^{(1)})^3\big]\,G_{t,\ell}(x,\xb)\\
	&~~~+\log^2(u)\,\big[(\tfrac{1}{2}A^{(1)}(\gamma^{(1)})^2+A^{(0)}\gamma^{(1)}\gamma^{(2)})+A^{(0)}(\gamma^{(1)})^3\partial_{\Delta}\big]\,G_{t,\ell}(x,\xb)\\
	&~~~+\log^1(u)\,\big[(A^{(2)}\gamma^{(1)}+A^{(1)}\gamma^{(2)}+A^{(0)}\gamma^{(3)})\\
	&\qquad\qquad\quad~~+2(A^{(1)}(\gamma^{(1)})^2+2A^{(0)}\gamma^{(1)}\gamma^{(2)})\partial_{\Delta}+2A^{(0)}(\gamma^{(1)})^3\partial_{\Delta}^2\big]\,G_{t,\ell}(x,\xb)\\
	&~~~+\log^0(u)\,\big[A^{(3)}+2(A^{(2)}\gamma^{(1)}+A^{(1)}\gamma^{(2)}+A^{(0)}\gamma^{(3)})\partial_{\Delta}\\
	&\qquad\qquad\quad~~+2(A^{(1)}(\gamma^{(1)})^2+2A^{(0)}\gamma^{(1)}\gamma^{(2)})\partial_{\Delta}^2+\tfrac{4}{3}A^{(0)}(\gamma^{(1)})^3\partial_{\Delta}^3\big]\,G_{t,\ell}(x,\xb)\,.\\
\end{split}
\end{align}
To facilitate readability, all indices and three kinds of summation have been suppressed: within each $\log(u)$-stratum, a finite sum over degeneracy labels $i=1,\ldots,t-1$, as well as two infinite sums over half-twists $t$ and even spins $\ell$.

Importantly, note that in the above we have only recorded the contributions from double-trace operators $\cO_{t,\ell,i}$. At twist 6 and above, triple-trace (and more generally, higher-trace) operators are expected to contribute and to mix with the double-trace spectrum. As we will show in Section~\ref{sec:triple-trace}, consistency of the OPE indeed predicts the presence of operators beyond just the double-trace spectrum in the $\log^2(u)$ part of $\mathcal{H}^{(3)}$ starting from twist 6.

\section{Zigzag integrals: a basis for the leading log}\label{sec:leading-logs}\setcounter{equation}{0}
When combining the perturbative expansion in $a$ with the superconformal block decomposition as shown for the first few orders in equation~\eqref{eq:logu_stratification}, one notices that the leading logarithmic divergence (or leading log for short) is entirely determined by the tree-level OPE data of long double-trace operators $\cO_{t,\ell,i}$. To be precise, at any order $a^n$, one has
\begin{align}\label{eq:leading_log}
	\mathcal{H}^{(n)}(u,v)|_{\log^n(u)} = \frac{1}{n!}\sum_{t,\ell}\sum_{i=1}^{t-1} A^{(0)}_{t,\ell,i} \big(\gamma^{(1)}_{i}\big)^n G_{t,\ell}(x,\xb),
\end{align}
which depends only on the leading-order OPE coefficients $A^{(0)}_{t,\ell,i}$ and tree-level anomalous dimensions $\gamma^{(1)}$ given in~\eqref{eq:A0} and~\eqref{eq:gamma1}, respectively. The simplicity of these suggest that a pattern will emerge in the resummed form of the leading log. As we will discuss in the following, this is indeed the case and moreover we can identify a family of four-dimensional loop integrals, the so-called `zigzag' integrals, which provide the correct functional basis for the leading log. This observation will later feed into our ansatz for the two-loop correlator $\mathcal{H}^{(3)}$, but here we make the more general observation that this structure is present at any loop-order.

\subsection{The operator $\d8$ and the leading log}\label{sec:delta8_and_leading-log}
To begin with, it is possible to directly perform the defining sums~\eqref{eq:leading_log} case by case and thus obtain explicit results for the leading log. These expressions are of the general form
\begin{align}
	\mathcal{H}^{(n)}(u,v)|_{\log^n(u)} = \frac{u^2~f_{\log}^{(n)}(x,\xb)}{(x-\xb)^{8n-1}},
\end{align}
where $f_{\log}^{(n)}(x,\xb)$ is given by a linear combination of HPL's of transcendental weights up to $n$ with polynomial coefficients of (combined) degree $12n-6$. However, as realised in~\cite{Aprile:2017qoy,Aprile:2018efk} and later explained in terms of a hidden ten-dimensional conformal symmetry~\cite{Caron-Huot:2018kta}, one can greatly simplify these expressions by making use of a certain eighth-order differential operator denoted by $\d8$. For the relevant case at hand, i.e. the $su(4)$ singlet representation, this operator can be written in the fully factorised form\footnote{
	While here we focus on the $\fourtwo$ correlator which contributes only to the singlet-channel, a generalisation of $\d8$ to all $su(4)$ channels $[a,b,a]$ (relevant for four-point correlators of arbitrary external charges) has been given in~\cite{Aprile:2018efk}. For a beautiful reformulation into one compact formula see~\cite{Caron-Huot:2018kta}.} \footnote{
	The properties of $\d8$ under crossing transformations are discussed in Section~\ref{sec:d8_symmetries}.}
\begin{align}\label{eq:d8_def}
	\d8 = \frac{u^4}{(x-\xb)}\partial_{\xb}^2(1-\xb)^2\partial_{\xb}^2 \partial_x^2(1-x)^2\partial_x^2(x-\xb).
\end{align}
The key observation made in~\cite{Aprile:2017qoy,Aprile:2018efk} is that $\d8$ satisfies an eigenvalue equation when acting on long superconformal blocks $u^2G_{t,\ell}$, with its eigenvalue being given (up to a factor of $-2$) by the numerator of the tree-level anomalous dimensions $\gamma^{(1)}$ from equation~\eqref{eq:gamma1}:
\begin{align}
	\d8\,u^2G_{t,\ell}(x,\xb) &= (t-1)_4(t+\ell)_4\,u^2G_{t,\ell}(x,\xb).
\end{align}
Since this numerator is independent of the degeneracy label $i$, one may pull out $n-1$ factors of $\d8$ from the sum~\eqref{eq:leading_log} and thus remove $n-1$ powers of the numerator from $(\gamma^{(1)})^n$, leading to a considerably simpler sum. Explicitly, this leads to
\begin{align}\label{eq:leading_log_d8}
	\mathcal{H}^{(n)}(u,v)|_{\log^n(u)} = \frac{1}{u^2}\big(\d8\big)^{n-1} g^{(n)}(x,\bar{x})\,, \qquad g^{(n)}(x,\bar{x}) = \frac{{u^2} \tilde{f}_{\log}^{(n)}(x,\xb)}{(x-\xb)^{7}},
\end{align}
where the functions $\tilde{f}_{\log}^{(n)}(x,\xb)$ contain the same HPL's as $f_{\log}^{(n)}(x,\xb)$ but their coefficient polynomials have greatly simplified now, with their degrees being reduced from $12n-6$ to only $10$.

A further simplification of the leading log can be achieved by making use of the observation that the ten-dimensional conformal blocks diagonalise the mixing of double-trace operators at order $a$ \cite{Caron-Huot:2018kta}. This is a consequence of the hidden conformal symmetry of tree-level supergravity correlators, effectively reducing the computation of the leading log to a single infinite sum over spins. We refer to Section 5.5 of~\cite{Caron-Huot:2018kta} for more details. Here we just quote the form of the result, which reduces the information required to a single-variable function $h^{(n)}(x)$,
\be
	h^{(n)}(x) = \frac{(-2)^{n+2}}{n!}\sum_{l=0,2,4,\ldots} \frac{960 \Gamma(l+1) \Gamma(l+4) x^{l+1} {}_2F_1(l+1,l+4;2l+8;x)}{\Gamma(2l+7) [(l+1)_6]^{n-1}}\,.
\ee
The function $h^{(n)}(x)$ is related to the function $g^{(n)}(x,\bar{x})$ appearing in (\ref{eq:leading_log_d8}) via a third-order differential operator,
\be
	g^{(n)}(x,\bar{x}) = \mathcal{D}_x^{(3)} h^{(n)}(x) + (x \leftrightarrow \bar{x})\,,
\ee
where $\mathcal{D}_x^{(3)}$ is given by
\be
	\mathcal{D}_x^{(3)} = -\biggl(\frac{x \bar{x}}{x-\bar{x}}\biggr)^7 + \frac{1}{2} \biggl(\frac{x \bar{x}}{x-\bar{x}}\biggr)^6 x^2 \partial_x - \frac{1}{10} \biggl(\frac{x \bar{x}}{x-\bar{x}}\biggr)^5 x^3 \partial_x^2 x + \frac{1}{120} \biggl(\frac{x \bar{x}}{x-\bar{x}}\biggr)^4 x^4 \partial_x^3 x^2\,.
\ee
Note that under the crossing transformation,
\be
	x \mapsto x' = \frac{x}{x-1}\,, \qquad \bar{x} \mapsto \bar{x}' = \frac{\bar{x}}{\bar{x}-1}\,,
\ee
the function $h^{(n)}$ and the operator $\mathcal{D}^{(3)}_x$ are both antisymmetric, 
\be
	h^{(n)}(x') = - h^{(n)}(x)\, \qquad  \mathcal{D}^{(3)}_x \mapsto \mathcal{D}^{(3)}_{x'} = - \mathcal{D}^{(3)}_x\,.
\ee

Some contributions to $h^{(n)}(x)$ (or $g^{(n)}(x,\xb)$) turn out to be so simple that closed form expressions have been found for them for any $n$. We define the function $J_k(x)$ as a linear combination of weight $k$ harmonic polylogarithms via
\be
	J_k(x) = \sum_{a_i=0,1} H_{0 a_1 0 a_2  \ldots 1}(x)\,, \qquad J_k(x') = -\sum_{a_i=0,1} H_{ a_1 0 a_2 0  \ldots 1}(x)\,.
\ee
Note that $J_1(x) = - J_1(x') = H_1(x)$. The function $h^{(n)}$ has the form
\be
	h^{(n)}(x) = a^{(n)}_0(x) + a^{(n)}_1(x) H_1(x) + \sum_{k=2}^n [ a^{(n)}_k(x) J_k(x) - (x \leftrightarrow x')]\,,
\ee
with $a^{(n)}_k(x)$ being polynomials in $1/x$ of degree at most six. The leading terms obey $a^{(n)}_0(x') = - a^{(n)}_0(x)$ and $a^{(n)}_1(x') = a^{(n)}_1(x)$ so that $h(x') = - h(x)$.
We see that $h^{(n)}(x)$ is organised by terms of transcendental weight $k=0,1,\ldots,n$. Since the functions $J_n$ for $n\geq2$ obey 
\be
	x \partial_x J_n(x) = - J_{n-1}(x')\, \qquad x' \partial_{x'} J_n(x') = x(1-x)\partial_x J_n(x') = - J_{n-1}(x)\,,
\ee
it follows that $g^{(n)}(x,\bar{x})$ is organised similarly,
\begin{align}\label{gweights}
\begin{split}
	g^{(n)}(x,\bar{x}) &= g^{(n)}_0(x,\bar{x}) + \bigl[g^{(n)}_1(x,\bar{x}) H_1(x) + (x \leftrightarrow \bar{x})\bigr] \\
 	&+\sum_{k=2}^{n} \bigl[ [g^{(n)}_k(x,\bar{x}) J_k(x) + g^{(n)}_k(x',\bar{x}') J_k(x')] + (x \leftrightarrow \bar{x})\bigr]\,.
\end{split}
\end{align}
Here $g^{(n)}_0$ obeys $g^{(n)}_0(x',\bar{x}') = g^{(n)}_0(x,\bar{x}) = g^{(n)}_0(\bar{x},x)$ and $g^{(n)}_1$ obeys $g^{(n)}_1(x',\bar{x}') = - g^{(n)}_1(x,\bar{x})$.

The terms of leading transcendental weight can be written for all $n$ \cite{Bissi:2020woe},
\be
	a^{(n)}_n(x) =  \frac{(-1)^n 4^{4-n} 15^{1-n}}{x^5 n!} \bigl((-6 \cdot 5^n + 10^n + 50)x^2 +3(5^n-25)x+30\bigr)\,.
\ee
This expression reproduces\footnote{up to an overall $n$-dependent prefactor due to differences in conventions.} the expression given in \cite{Aprile:2018efk} for the highest weight contribution (i.e. the $k=n$ term in (\ref{gweights})) to $g^{(n)}(x,\bar{x})$,
\be
\label{topweightg}
	g^{(n)}_{\rm top}(x,\bar{x}) = \bigl[J_n(x) \mathcal{D}_x^{(3)} a^{(n)}_n(x)  -  J_n(x') \mathcal{D}_x^{(3)} a^{(n)}_n(x')\bigr] + (x \leftrightarrow \bar{x})\,.
\ee
Note that when the operator $\mathcal{D}^{(3)}_x$ acts on a polynomial in $1/x$ of degree at most six, the result is always antisymmetric under $(x \leftrightarrow \bar{x})$. For the leading weight term (\ref{topweightg}) it follows that the explicit symmetrisation actually produces antisymmetric combinations of harmonic polylogs of the form $J_n(x) - J_n(\bar{x})$ and $J_n(x') - J_x(\bar{x}')$,
\be
	g^{(n)}_{\rm top}(x,\bar{x}) = [J_n(x)-J_n(\bar{x})] \mathcal{D}_x^{(3)} a^{(n)}_n(x)  -  [J_n(x')-J_n(\bar{x}')] \mathcal{D}_x^{(3)} a^{(n)}_n(x') \,.
\label{gtopsym}
\ee
For the lower weight terms ($k<n$ in (\ref{gweights})) we generically have different coefficients for $J_k(x)$, $J_k(\bar{x})$, $J_k(x')$ and $J_k(\bar{x}')$. Note that a closed form has also been found for $a^{(n)}_{n-1}$ in~\cite{Bissi:2020woe}.

While these formulae are very explicit and a clear pattern is visible in the HPL's which do appear, the origins of these structures were left unexplained. In the following, we describe a family of four-dimensional loop integrals which give rise to these previously observed structures. 

\subsection{The Zigzag integrals}\label{sec:zigzag}
The relevant series are the zigzag series of integrals which obey
\be\label{zigzagdiffeq}
	x\xb \partial_x \partial_{\xb} Z^{(L)}(x,\xb) = Z^{(L-1)}(1-x,1-\xb)\,.
\ee
where the series begins with the one-loop ladder function
\be
	Z^{(1)}(x,\bar{x}) = \phi^{(1)}(x,\bar{x})\,,
\ee
and the solution for all higher $L$ is determined uniquely by imposing single-valuedness. Because of the extra symmetry of $\phi^{(1)}$ we find $Z^{(2)} = \phi^{(2)}$ but that thereafter the series is different from the ladder series $\phi^{(L)}$~\cite{Usyukina:1993ch}. The functions $Z^{(n)}$ are antisymmetric under~$x \leftrightarrow \bar{x}$.

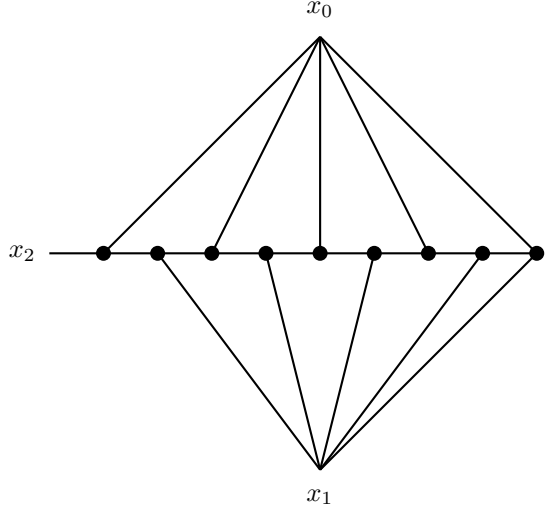
\begin{figure}
	{\footnotesize
		\begin{center}
			\begin{tikzpicture}[scale=0.36]
			\draw[join=bevel,thick] (-10,0) -- (8,0);
			\draw[join=bevel,thick] (0,8) -- (8,0);
			\draw[join=bevel,thick] (0,-8) -- (8,0);
			\draw[join=bevel,thick] (-8,0) -- (0,8);
			\draw[join=bevel,thick] (-6,0) -- (0,-8);
			\draw[join=bevel,thick] (-4,0) -- (0,8);
			\draw[join=bevel,thick] (-2,0) -- (0,-8);
			\draw[join=bevel,thick] (0,0) -- (0,8);
			\draw[join=bevel,thick] (2,0) -- (0,-8);
			\draw[join=bevel,thick] (4,0) -- (0,8);
			\draw[join=bevel,thick] (6,0) -- (0,-8);
			\node[circle,fill,inner sep=2pt] at (8,0){};
			\node[circle,fill,inner sep=2pt] at (6,0){};
			\node[circle,fill,inner sep=2pt] at (4,0){};
			\node[circle,fill,inner sep=2pt] at (2,0){};
			\node[circle,fill,inner sep=2pt] at (0,0){};
			\node[circle,fill,inner sep=2pt] at (-2,0){};
			\node[circle,fill,inner sep=2pt] at (-4,0){};
			\node[circle,fill,inner sep=2pt] at (-6,0){};
			\node[circle,fill,inner sep=2pt] at (-8,0){};
			\node at (-11,0){$x_2$};
			\node at (0,9){$x_0$};
			\node at (0,-9){$x_1$};
			\end{tikzpicture}
		\end{center}
	}
	\caption{Example of a zigzag integral, with integration vertices marked with dots.}
	\label{Fig:Zigzag}
\end{figure}

Both the ladder series and the zigzag series arise from a more general class of four-dimensional loop integrals \cite{Drummond:2012bg}. Given a word $m = a_1 ... a_{L-1}$ with letters $a_i \in \{0,1\}$, we define an $L$-loop, three-point  integral as follows,
\be
	I_m(x_0,x_1,x_2) = \frac{1}{\pi^{2L}}\int \frac{1}{x_{2 b_1}^2} \prod_{i=1}^{L-1} \biggl(\frac{d^4x_{b_i}}{x_{b_i b_{i+1}}^2 x_{b_i a_i}^2}\biggr) \frac{d^4x_{b_L}}{x_{b_L 0}^2 x_{b_L 1}^2}\,.
\ee
The integration vertices in the above definition are labelled by $x_{b_i}$ for $i=1,\ldots,L$ and we use the shorthand $x_{ij}^2 = (x_i-x_j)^2$. If all the $a_i$ are zero then the above integral $I_m$ corresponds to the three-point version of the dual of the ladder series.\footnote{
	I.e. the limit as the fourth point is taken to infinity of the conformal four-point integral dual to the $L$-loop ladder diagram.}
If we alternate the indices $m = a_1 a_2 \ldots a_{L-1} = 0101\ldots$ we get the zigzag series (with an example illustrated in Fig.~\ref{Fig:Zigzag}),
\be
\label{zigzagintfn}
	I_{m}(x_0,x_1,x_2) = \frac{1}{x_{01}^2} \frac{Z^{(L)}(z,\bar{z})}{z-\bar{z}}\,.
\ee
Here we have used the variables,
\be
	\frac{z \bar{z}}{(1-z)(1-\bar{z})} = \frac{x_{02}^2}{x_{12}^2}\,, \qquad \frac{1}{(1-z)(1-\bar{z})} = \frac{x_{01}^2}{x_{12}^2}\,.
\ee
Note that the zigzag function is antisymmetric, $Z^{(n)}(\bar{z},z) = - Z^{(n)}(z,\bar{z})$.

The differential equation (\ref{zigzagdiffeq}) follows from considering the action of the Laplace operator at the point $x_2$ on both sides of (\ref{zigzagintfn}). Since the Laplace operator acts on a single propagator in the integral, its action produces a delta-function which localises one loop integral to produce an $(L-1)$ loop integral, which is also a zigzag after swapping the points $x_0$ and $x_1$. The differential equation together with the constraint that the function should be single-valued (in Euclidean kinematics where $\bar{z}$ is the complex conjugate of $z$) uniquely defines $Z^{(n)}$.

Here we give expressions for the first few zigzags $Z^{(n)}$ in terms of single-valued polylogarithms \cite{fbsvpl},
\begin{align}
\begin{split}
	Z^{(1)} &= \mathcal{L}_{2} - \mathcal{L}_{10}\,,\\
	Z^{(2)} &= \mathcal{L}_{200} - \mathcal{L}_{30} \,,\\
	Z^{(3)} &= \mathcal{L}_{2210} - \mathcal{L}_{2120} - 2 \zeta_3(3\mathcal{L}_{20} + 2 \mathcal{L}_{21})\,,\\
	Z^{(4)} &= \mathcal{L}_{2230} - \mathcal{L}_{2320} - 4 \zeta_3(\mathcal{L}_{23} -  \mathcal{L}_{220}) - 20 \zeta_5 \mathcal{L}_{20}\,,\\
	Z^{(5)} &= \mathcal{L}_{222120} - \mathcal{L}_{221220} + 4 \zeta_3(\mathcal{L}_{2221} - \mathcal{L}_{2212}) + \zeta_5(4\mathcal{L}_{221} +15 \mathcal{L}_{220})\\
	&\quad - 12 \zeta_{3}^2 \mathcal{L}_{22} - \frac{441}{8} \zeta_7 \mathcal{L}_{20} + 18 \zeta_3 \zeta_5 \mathcal{L}_2\,.
\end{split}
\end{align}
It is worth remarking that the detailed form of the zeta-values appearing above is directly related to the values of certain vacuum integrals (also called zigzags), whose name motivates our name for the $Z^{(n)}$, and whose form was conjectured to all loops in \cite{Broadhurst:1995km} and proven in~\cite{Brown:2015ztw}.\footnote{
	See also the recent work~\cite{Derkachov:2022ytx}.}

Let us note that one may take many $\partial_x$ derivatives of the zigzag functions and obtain other single-valued pure transcendental functions. In fact one may take $m$ derivatives with $m\leq (n-1)$, alternating between $(-x\partial_x)$ and $(1-x)\partial_x$ and obtain another pure single-valued function of weight $2n-m$,
\be\label{eq:zigzag_derivatives}
	Z_m^{(n)}(x,\bar{x}) = \underbrace{\ldots (1-x)\partial_x (-x \partial_x) (1-x)\partial_x (-x \partial_x)}_{m \text{ derivatives}} Z^{(n)}(x,\bar{x})\,.
\ee
If we take a further derivative we obtain a sum of two pure functions of weight $n$ with different rational prefactors. Obviously, the functions $Z_m^{(n)}(x,\bar{x})$ do not exhibit any symmetry under $x \leftrightarrow \bar{x}$ although they may be decomposed into symmetric and antisymmetric parts.

None of the zeta value terms contribute to the leading discontinuity which is therefore simple to calculate. Indeed the leading logarithmic singularities of the zigzags obey
\begin{align}
\begin{split}\label{zigzagsleadinglog}
	Z^{(n)}\biggl(\frac{1}{x},\frac{1}{\bar{x}}\biggr)\bigg|_{\log^n(u)} &= \frac{(-1)^{n+1}}{n!} [J_n(x)-J_n(\bar{x})]\,,\\
	Z^{(n)}\biggl(\frac{1}{x'},\frac{1}{\bar{x}'}\biggr)\bigg|_{\log^n(u)} &= \frac{(-1)^{n+1}}{n!} [J_n(x')-J_n(\bar{x}')]\,.
\end{split}
\end{align}
These leading discontinuities are exactly what is required to match the leading weight terms predicted by OPE, as given in (\ref{gtopsym}). More precisely, to match the top weight terms in the leading discontinuity we should include in $\mathcal{H}^{(n)}(u,v)$ the contribution
\be
	\frac{1}{u^2}\bigl(\Delta^{(8)}\bigr)^{n-1} \mathcal{G}^{(n)}_{\rm top}(x,\bar{x})\,,
\ee
where
\be
\label{Ztop}
\mathcal{G}^{(n)}_{\rm top}(x,\bar{x}) = (-1)^{n+1} n! \bigl[\mathcal{D}_x^{(3)} a^{(n)}_n(x)\bigr] Z^{(n)}\biggl(\frac{1}{x},\frac{1}{\bar{x}}\biggr) + (x \leftrightarrow x') \,.
\ee
Recall that $\mathcal{D}_x^{(3)} a^{(n)}_n(x)$ is actually an antisymmetric function of $x$ and $\bar{x}$, as is $Z^{(n)}$, so that $\mathcal{G}^{(n)}_{\rm top}$ is symmetric.

Moreover we can also see that taking $\partial_x$ derivatives of the orientations of the zigzags given in (\ref{zigzagsleadinglog}) will give single-valued functions whose leading discontinuities will give the contributions to $g^{(n)}(x,\bar{x})$ of the form $J_k(x)$ and $J_k(x')$ for $k<n$. For $1\leq m \leq n-1$ we have
\begin{align}
\begin{split}
	Z_m^{(n)}\biggl(\frac{1}{x},\frac{1}{\bar{x}}\biggr)\bigg|_{\log^n(u)} & =
	\begin{cases}
	\frac{(-1)^{n-m+1}}{n!} J_{n-m}(x')\,, \text{ $m$ odd}  \\
	\frac{(-1)^{n-m+1}}{n!} J_{n-m}(x)\,, \text{ $m$ even} 
	\end{cases} \\
	Z_m^{(n)}\biggl(\frac{1}{x'},\frac{1}{\bar{x}'}\biggr)\bigg|_{\log^n(u)} &= 
	\begin{cases}
	\frac{(-1)^{n-m+1}}{n!} J_{n-m}(x)\,, \text{ $m$ odd}  \\
	\frac{(-1)^{n-m+1}}{n!} J_{n-m}(x')\,, \text{ $m$ even} 
	\end{cases}
\end{split}
\end{align}
Since we have acted with $\partial_x$ derivatives the contributions of the form $J_n(\bar{x})$ from (\ref{zigzagsleadinglog}) have disappeared. By replacing the $J_k(x)$ and $J_k(x')$ and conjugates in $g^{(n)}(x,\bar{x})$ by the appropriate zigzag derivative, one may therefore construct a function which matches the leading discontinuity to all weights from the zigzags $Z^{(n)}$ and their derivatives. Let us consider the following combination,
\begin{align}\label{cG}
	\mathcal{G}^{(n)}(x,\bar{x}) = 
	&  \Biggl[\frac{n!}{4}\bigl( g^{(n)}_0(x,\bar{x})(1-x)\partial_x + 2 g^{(n)}_1(x,\bar{x})\bigr) \Bigl( Z^{(n)}_{n-1}\biggl(\frac{1}{x},\frac{1}{\bar{x}}\biggr) - (x \leftrightarrow x') \Bigr) + (x \leftrightarrow \bar{x})\Biggr]\notag\\
	&-n! \sum_{k=2}^{n-1} \Bigl[ \bigl( (-1)^k g^{(n)}_k(x,\bar{x}) Z^{(n)}_{k-n}(z_k) + (x \leftrightarrow x')\bigr) + (x \leftrightarrow \bar{x})\Bigr]\notag\\
	&+ \mathcal{G}^{(n)}_{\rm top}(x,\bar{x})\,.
\end{align}
where $z_k = 1/x$ for $(n+k)$ even and $z_k=1/x'$ for $(n+k)$ odd.

If we take the leading discontinuity of this function we obtain $g^{(n)}(x,\bar{x})$. 
We may then include a contribution to $\mathcal{H}^{(n)}(u,v)$ as follows,
\be\label{eq:leading_log_manifest}
	\mathcal{H}^{(n)}(u,v) = \frac{1}{u^2} \bigl(\Delta^{(8)}\bigr)^{n-1} \mathcal{G}^{(n)}(x,\bar{x}) + \text{ terms with no leading discontinuity}\,.
\ee

For $n=1,2$ the zigzags are related to ladder integrals, as we have seen, and hence for these values they have additional symmetries under crossing transformations. For $n\geq 3$ the integrals do not exhibit any symmetries under crossing transformations and the six possible crossing orientations which we may list as,
\be
Z^{(n)}(x,\bar{x}), \, Z^{(n)}(1-x,1-\bar{x}), \, Z^{(n)}\biggl(\frac{1}{x'},\frac{1}{\bar{x}'}\biggr), \, Z^{(n)}(x',\bar{x}'), \, Z^{(n)} \biggl(\frac{1}{1-x},\frac{1}{1-\bar{x}}\biggr), \, Z^{(n)}\biggl(\frac{1}{x},\frac{1}{\bar{x}}\biggr),
\notag
\ee
are linearly independent. Moreover, for $n\geq 3$ the only orientations which contribute to the leading log term are those given in (\ref{zigzagsleadinglog}). This makes matching the leading logarithmic discontinuities with a combination of all orientations of zigzag integrals (which will be required for crossing invariance of the final answer) straightforward at two loops and beyond.\\

In summary, we have seen how the leading discontinuity at any loop order admits a compact expression with powers of the differential operator $\d8$ pulled out. Moreover, we have shown that the transcendental structure of the leading log is described solely in terms of zigzag integrals and derivatives thereof. As depicted schematically in equation \eqref{eq:leading_log_manifest}, the remaining difficulty in determining the full correlator $\mathcal{H}^{(n)}$ then lies in the construction of the remaining terms with no contribution to the leading log, which in general will include transcendental functions beyond the zigzag integrals discussed here. We will address this task for the two-loop case in Section \ref{sec:two_loops}.

\section{The bulk-point limit of AdS$_5\times$S$^5$ amplitudes}\label{sec:bulk_point}\setcounter{equation}{0}
Let us next discuss the flat-space limit of supergraviton amplitudes on AdS$_5\times$S$^5$ as described by the dual CFT correlator $\mathcal{H}(x,\xb)$. In its position space representation considered here\footnote{%
	A similar formulation of the flat-space limit for the Mellin space representation of holographic correlators has been developed in~\cite{Penedones:2010ue,Fitzpatrick:2011hu}. This flat-space limit of Mellin amplitudes has proven to be a particularly useful tool for constraining tree-level string corrections in AdS, see e.g. references~\cite{Goncalves:2014ffa,Alday:2018pdi,Binder:2019jwn,Drummond:2019odu,Drummond:2020dwr,Aprile:2020mus}. A beautiful formalism adapted specifically to the AdS$_5\times$S$^5$ case has been proposed in~\cite{Aprile:2020luw}, and for discussions of the flat-space limit for one-loop Mellin amplitudes see references~\cite{Alday:2018kkw,Alday:2019nin,Alday:2020tgi,Alday:2021ajh}.}
this amounts to considering the so-called bulk-point limit: By using sufficiently localised wave-packets in AdS, one can focus onto a point in the bulk and thus effectively recover the corresponding scattering process in flat-space. This manifests itself in a singularity of the correlator $\mathcal{G}(x,\xb;y,\bar{y})$ which is of the form
\begin{align}
	\mathcal{H}(x,\xb)~\xrightarrow[]{~\text{bpl}~}~\frac{\mathcal{F}(x)}{(x-\xb)^{p}}\,,
\end{align}
for some positive integer power $p$ specified later. As explained in references~\cite{Gary:2009ae,Heemskerk:2009pn,Okuda:2010ym,Maldacena:2015iua}, this is the expected behaviour of holographic correlators with a local bulk dual. Note that in order to expose this bulk-point singularity, one first needs to analytically continue the Euclidean correlator to Lorentzian signature before taking the limit $\xb\to x$. The residue of the singularity is then directly related to the corresponding (in this case ten-dimensional) flat-space scattering amplitude $\ca^{(10)}$ as a function of the scattering angle,
\begin{align}\label{eq:bulk-point_relation}
	\ca^{(10)}\,\propto\,s^k\,\frac{\mathcal{F}(x)}{x^{l}(1-x)^{m}}\,,
\end{align}
where $k,l,m$ are integers related to the dimension of the effective bulk interaction vertex~\cite{Heemskerk:2009pn}. The dimensionless parameter $x$ is defined in terms of the Mandelstam invariants $s$ and $t$, or equivalently in terms of the scattering angle $\theta$, by
\begin{align}\label{eq:scattering_angle}
	x\equiv 1+\frac{t}{s}=\frac{1+\cos\theta}{2}.
\end{align}
For our purposes, the relevant flat-space amplitude $\ca^{(10)}$ is given by the massless four-particle scattering amplitude of ten-dimensional type IIB string theory, whose low-energy expansion we will review in the next section.

The above relation~\eqref{eq:bulk-point_relation} can be leveraged to put constraints on the AdS correlator by using knowledge about the corresponding flat-space amplitude. Let us first point out that in the context of the large $N$ expansion~\eqref{eq:large_N_expansion} of the $\mathcal{N}=4$ SYM correlator, in~\cite{Alday:2017vkk} a simpler matching than given by~\eqref{eq:bulk-point_relation} has been proposed, namely a relation between the double-discontinuity $\text{dDisc}\,\mathcal{H}(x,\xb)$ and the $t$-channel discontinuity $\text{Disc}_t\,\ca^{(10)}(x)$. Therein, the authors applied it to the one-loop correlator $\mathcal{H}^{(2)}$, with later applications to one-loop string corrections given in~\cite{Alday:2018pdi}. At two-loop order, this simpler matching has been considered for the leading log of $\mathcal{H}^{(3)}$ in the works~\cite{Bissi:2020wtv,Bissi:2020woe}, and also the recently proposed two-loop bootstrap result of~\cite{Huang:2021xws} imposes this simpler matching.

In contrast, here we will describe how in the bulk-point limit the large $N$, large $\l$ expansion of $\mathcal{H}(x,\xb)$ recovers the corresponding low-energy expansion of the \textit{full} type IIB string amplitude, not just its discontinuity. In general, one would not expect this to lead to more constraints for the AdS correlator than the simpler matching performed at the level of the discontinuities, since the Lorentzian inversion formula of~\cite{Caron-Huot:2017vep} allows one to (in principle) reconstruct the full correlator from its double-discontinuity.\footnote{
	In analogy with the reconstruction of flat-space amplitudes from their discontinuity up to regular terms.
	}
However, this statement holds only modulo terms with vanishing double-discontinuity and finite spin contributions in the conformal block decomposition. But this is exactly the case for the tree-level string corrections $\mathcal{H}^{(1,m)}$ with $m>0$, which subsequently also appear at loop-order in the form of finite-spin ambiguities.

Hence, besides providing an explicit test of the relation~\eqref{eq:bulk-point_relation} across different orders in the perturbative expansion, the extended matching proposed here is sensitive to tree-level contributions and finite-spin ambiguities of loop amplitudes in AdS. Unfortunately, as we will explain in the following, some of the terms in the low-energy expansion of the genus-one string amplitude depend on some constant scale $\mu$. This prevents us from fixing the tree-level ambiguities in the one-loop string corrections $\mathcal{H}^{(2,m)}$ which contribute to the bulk-point limit, since shifting the scale alters the value of the non-analytic terms one would like to match. Nevertheless, the bulk-point limit of the one-loop correlators $\mathcal{H}^{(2,m)}$ sheds light on the role of functions containing the letter $x-\xb$. In particular, we will argue that the extra logarithmic divergence due to the presence of the letter $x-\xb$ is necessary to recover the scale-dependent logarithmic terms of the form $\log(\frac{\ap s}{\mu})$, which do appear in the low-energy expansion of the type IIB string amplitude starting from genus-one.

Lastly, moving to genus-two, we will describe a constraint for the bulk-point limit of the two-loop correlator $\mathcal{H}^{(3)}$ which comes from matching the two-loop supergravity amplitude in ten-dimensional flat-space.

\subsection{The perturbative expansion of the type IIB string amplitude}
Let us begin by assembling the necessary terms in the low-energy expansion of the type IIB string theory scattering amplitude $\ca^{(10)}$. Its genus-expansion takes the general form
\begin{align}\label{eq:flat_space_amplitude}
	\ca^{(10)} &= \kappa_{10}^2\,g_s^2\,\frac{\widehat{K}}{64}\,\big(\ca_{\text{genus-0}} + g_s^2\,\ca_{\text{genus-1}} + g_s^4\,\ca_{\text{genus-2}}+O(g_s^6)\big),
\end{align}
where $\kappa_{10}^2=2^6\pi^7\,(\ap)^4$ and $\widehat{K}$ is an overall kinematic factor depending on 10-dimensional momenta $k_i$ and polarisation vectors $\xi_i$ of the external gravitons. In order to compare with the bulk-point limit of holographic correlators on the product space AdS$_5\times$S$^5$, we need to restrict $\ca^{(10)}$ to transverse kinematics by taking the momenta of the external gravitons to lie in AdS$_5$ and their polarisation vectors along S$^5$, such that $k_i\cdot\xi_j=0$. As shown in~\cite{Chester:2018dga}, with such kinematics the overall factor $\widehat{K}\to\widehat{K}_{\perp}$ becomes proportional to the prefactor $\mathcal{I}$ of the dynamical part dictated by supersymmetry, recall equations~\eqref{eq:partial_non-ren} and~\eqref{eq:intriligator_factor}, and hence these two factors will cancel in the comparison~\eqref{eq:bulk-point_relation}. Moreover, we need to express the flat-space amplitude as a function of the dimensionless parameter $x=1+\frac{t}{s}$, related to the scattering angle via~\eqref{eq:scattering_angle}. This is achieved by pulling out the overall scaling with the center-of-mass energy $s$ from the individual terms in the expansion~\eqref{eq:flat_space_amplitude}. We thus rewrite the low-energy expansion in the form
\begin{align}
	\ca^{(10)}_{\perp}=\kappa_{10}^2\,g_s^2\,\frac{\widehat{K}_{\perp}}{64}\,\Big(\frac{1}{(\ap s)^3}\,\ca^{(0,-3)}(x)\,+\,\sum_{g=0}\sum_{k=0}(2\pi^2 g_s^2)^g(\ap s)^k\,\ca^{(g,k)}(x)\Big),
\end{align}
where $\ca^{(10)}_{\perp}$ denotes the 10-dimensional amplitude in transverse kinematics. The first term is the tree-level supergravity contribution which scales as $1/s^3$, followed by an infinite tower of string and higher-genus corrections, all scaling with positive powers of $s$.

The genus-zero contributions are given by the well-known Virasoro-Shapiro amplitude, and for the first few orders in $\ap$ one finds
\begin{align}\label{eq:genus0-terms}
\begin{split}
	\ca^{(0,-3)}&=\frac{64}{x(1-x)},\quad\ca^{(0,0)}=2\zeta_3,\quad\ca^{(0,2)}=\frac{\zeta_5}{8}(x^2-x+1),\quad\ca^{(0,3)}=\frac{\zeta_3^2}{2^{5}}x(1-x),\\[3pt]
	\ca^{(0,4)}&=\frac{\zeta_7}{2^7}(x^2-x+1)^2,\quad\ca^{(0,5)}=\frac{\zeta_3\zeta_5}{2^8}x(1-x)(x^2-x+1),\\[3pt]
	\ca^{(0,6)}&=\frac{\zeta_3^3}{2^{10}\,3}x^2(1-x)^2 + \frac{\zeta_9}{2^{11}\,3}(3x^6-9x^5+19x^4-23x^3+19x^2-9x+3),
\end{split}
\end{align}
where the terms $\ca^{(0,k)}$ with $k\geq0$ correspond to tree-level $\partial^{2k}\mathcal{R}^4$ higher-derivative corrections to the supergravity effective action.

Next, the low-energy expansion of the genus-one contribution $\ca_{\text{genus-1}}$ has been worked out most systematically in~\cite{Green:2008uj}.\footnote{See also the corrections/extensions given in references~\cite{DHoker:2015gmr,DHoker:2019blr}.} Putting together the analytic and non-analytic contributions, we have
\begin{align}\label{eq:genus1-terms}
\begin{split}
	\ca^{(1,0)}&=\frac{1}{3}\,,\quad\ca^{(1,1)}=\ca^\text{one-loop}_{\text{sugra}}(x)\,,\quad\ca^{(1,2)}=0\,,\quad\ca^{(1,3)}=\frac{\zeta_3}{2^{6}\,3}x(1-x)\,,\\[3pt]
	\ca^{(1,4)}&=-\frac{\zeta_3}{2^{6}\,45}\Big[x^4\log(x)+(1-x)^4\log(1-x)-i\pi+2(x^2-x+1)^2 \log(\tfrac{\ap\,s}{\mu_4})\Big]\,,\\[3pt]
	\ca^{(1,5)}&=\frac{29\,\zeta_5}{2^{11}\,45}x(1-x)(x^2-x+1)\,,\\[3pt]
	\ca^{(1,6)}&=\frac{\zeta_3^2}{2^{13}\,3}x^2(1-x)^2-\frac{\zeta_5}{2^{12}\,315}\Big[x^4(22x^2-x+1)\log(x)\\
				&\qquad\qquad\quad+(1-x)^4(22x^2-43x+22)\log(1-x)-i\pi(x^2-x+22)\\
				&\qquad\qquad\quad+(44x^6-132x^5+327x^4-434x^3+327x^2-132x+44)\log(\tfrac{\ap\,s}{\mu_6})\Big]\,,
\end{split}
\end{align}
for some dimensionless scales $\mu_4$ and $\mu_6$, which are determined by the perturbative string theory expansion. Note that the contribution to $\ca^{(1,1)}$ is given by the one-loop supergravity field theory amplitude $\ca^\text{one-loop}_{\text{sugra}}(x)$ which scales linearly with $s$.

Lastly, we move to the genus-two term $\ca_{\text{genus-2}}$ where much less is known about its low-energy expansion. To our knowledge, so far only the first two non-trivial analytic terms, corresponding to the genus-two corrections of the $\partial^4\mathcal{R}^4$ and $\partial^6\mathcal{R}^4$ terms, have been determined~\cite{DHoker:2005jhf,Gomez:2010ad,DHoker:2014oxd}:
\begin{align}\label{eq:genus2-terms_1}
	\ca^{(2,0)}=\ca^{(2,1)}=0\,,\quad\ca^{(2,2)}=\frac{1}{2^{4}\,135}(x^2-x+1)\,,\quad\ca^{(2,3)}=\frac{1}{2^{7}\,15}x(1-x)\,,
\end{align}
The precise form of the terms beyond $\partial^6\mathcal{R}^4$ is not known, and the only information about the structure of these super-leading counter-terms to the two-loop supergravity amplitude comes from studying the compactification of 11-dimensional two-loop supergravity~\cite{Green:2008bf}. From those considerations, the next two terms in the expansion are expected to be of the schematic form
\begin{align}\label{eq:genus2-terms_2}
	\ca^{(2,4)}&\simeq\partial^{8}\mathcal{R}^4\vert_{\text{genus-2}}+\ca_{\text{SG}\vert\mathcal{R}^4\vert_{\text{genus-1}}}^{\text{one-loop}}\,,\quad\ca^{(2,5)}\simeq\partial^{10}\mathcal{R}^4\vert_{\text{genus-2}}+\ca_{\text{sugra}}^{\text{two-loops}}\,,
\end{align}
where $\ca_{\text{SG}\vert\mathcal{R}^4\vert_{\text{genus-1}}}^{\text{one-loop}}$ is a non-analytic contribution similar to the genus-one term $\ca^{(1,4)}$ from equation~\eqref{eq:genus1-terms}, arising due to a one-loop diagram with one supergravity and one $\mathcal{R}^4$-vertex. Finally, the relevant term for us is the contribution $\ca^{(2,5)}$, which is given by the sum of the genus-two correction to the tree-level $\partial^{10}\mathcal{R}^4$ term (whose precise coefficient is not known) and the ten-dimensional two-loop supergravity amplitude $\ca_{\text{sugra}}^{\text{two-loops}}$.

\subsection{The contributions $\mathcal{H}^{(n,m)}$ in the bulk-point limit}
As mentioned earlier, the bulk-point singularity of the CFT correlator can be accessed only in the Lorentzian regime.
The necessary analytic continuation from the Euclidean correlator $\mathcal{H}(x,\xb)$ corresponds to a Wick rotation of AdS global time, see e.g.~\cite{Gary:2009ae}. In terms of cross-ratios, the prescription is to take $x$ counter-clock wise around $0$ and $\xb$ around $1$. Then, the bulk-point singularity is exposed by taking the limit $\xb\to x$.
We find that the terms $\mathcal{H}^{(n,m)}$ at order $a^n\l^{-\frac{m}{2}}$ in the large $N$, large $\l$ expansion around the supergravity limit diverge as
\begin{align}\label{eq:bulk-point_relation2}
	\mathcal{H}^{(n,m)}(x,\xb)~\xrightarrow[]{~\text{bpl}~}~ \frac{\mathcal{F}^{(n,m)}(x)}{(x-\xb)^{p}}\,,
\end{align}
with the denominator power given by $p=8n+2m-1$.

For terms $\mathcal{H}^{(n,m)}$ given entirely in terms of SVHPL's, which is the case for all tree-level correlators $\mathcal{H}^{(1,m)}$ as well as the one-loop supergravity correction $\mathcal{H}^{(2,0)}$, taking the bulk-point limit as described above is relatively straightforward. For example, performing the analytic continuation and taking the limit $\xb\to x$ for the first few zigzag integrals, one finds
\begin{align}\label{eq:bpl_zigzags}
\begin{split}
	Z^{(1)}(x,\xb)~&\xrightarrow[]{~\text{bpl}~}~4\pi^2,\\
	Z^{(2)}(x,\xb)~&\xrightarrow[]{~\text{bpl}~}~-4\pi^2\big(H_{0,0}(x)+i\pi H_0(x)\big),\\
	Z^{(3)}(x,\xb)~&\xrightarrow[]{~\text{bpl}~}~8\pi^2\big(H_{3,1}(x)-i\pi(H_3(x)-\zeta_3)+\tfrac{1}{6}\pi^2 H_2(x)-\tfrac{11}{360}\pi^4\big).
\end{split}
\end{align} 

Then, the last step before comparing with the terms $\ca^{(g,k)}$ from the flat-space string amplitude is to convert the $\mathcal{N}=4$ SYM double-expansion in $a$ and $1/\l$ to string theory quantities $g_s$ and $\ap$. Using the usual AdS/CFT relations, we have
\begin{align}\label{eq:dictionary}
	a=\frac{1}{N^2-1}~\simeq~\frac{16\pi^2g_s^2\,(\ap)^4}{L^8}\,,\qquad \l^{-\frac{1}{2}}~\simeq~\frac{\ap}{L^2}\,,
\end{align}
where $L$ is the AdS radius which we set to one. The first relation in~\eqref{eq:dictionary} gives that increasing the loop-order in AdS by one corresponds to four powers of $\ap$, and hence the residues $\mathcal{F}^{(n,m)}(x)$ from~\eqref{eq:bulk-point_relation2} should be compared with the flat-space contributions $\ca^{(g,k)}$ with $g=n-1$ and $k=4n+m-7$.

\subsubsection*{Matching the genus-zero terms}
We begin by comparing against the genus-zero terms $\ca^{(0,k)}$ given in~\eqref{eq:genus0-terms}, whose AdS counterparts $\mathcal{H}^{(1,m)}$ have been considered up to order $(\ap)^9$ using the Mellin space representation.\footnote{
	Here, we use the results of~\cite{Aprile:2020mus}, see also~\cite{Abl:2020dbx} for a manifestly 10-dimensional formulation.
	}
Note that the comparison at tree-level is in some sense tautological, since precisely the terms which contribute to the bulk-point limit were fixed using the flat-space limit formulation in Mellin space in the first place. Nevertheless, this serves as a consistency check and enables us to fix some factors and the overall normalisation.

Converting the Mellin space results to their position space representation in terms of linear combinations of $\dbar{}$-functions and taking the bulk-point limit using the first line of~\eqref{eq:bpl_zigzags}, we find that the precise matching reads\footnote{
	As explained below~\eqref{eq:flat_space_amplitude}, the overall kinematic factor $\widehat{K}$ in transverse kinematics cancels against the factor $\mathcal{I}$. This cancellation is implicit in equation~\eqref{eq:bulk-point_matching}.
	}
\begin{align}\label{eq:bulk-point_matching}
	\ca^{(g=n-1,k)}(x) &= \frac{1}{2^{2k-3n+9}\,(2k+11)!}\frac{\mathcal{F}^{(n,m)}(x)}{x^{k+10} (x-1)^{k+6}}\,,
\end{align}
where we recall that $k=4n+m-7$. The above relation between the bulk-point limit of the correlators $\mathcal{H}^{(n,m)}$ and the corresponding terms in the flat-space amplitude is in agreement with the general expectation previously given in equation~\eqref{eq:bulk-point_relation}.

\subsubsection*{Matching the genus-one terms}
At genus-one, the coefficient of the super-leading counter-term $\mathcal{H}^{(2,-1)}$ is fixed from the flat-space limit alone~\cite{Chester:2019pvm} and hence it matches the analytic genus-one contribution $\ca^{(1,0)}$ by construction, in analogy with the tree-level terms discussed before.

More interesting is the comparison with the non-analytic terms. The first such contribution occurs at order $\ap$ and is the one-loop supergravity amplitude $\ca^\text{one-loop}_{\text{sugra}}(x)$, which can be determined from a field-theory computation as a sum over three orientations of the massless one-loop box-integral in 10 dimensions, see e.g. Appendix D of~\cite{Alday:2017vkk} for an explicit evaluation. Taking the bulk-point limit of the supergravity correlator $\mathcal{H}^{(2)}$ from~\cite{Aprile:2017bgs},\footnote{
	From the basis of 15 independent SVHPL's appearing in $\mathcal{H}^{(2)}$, only 6 elements contribute to the bulk-point limit. These are the three orientations of the two-loop zigzag $Z^{(2)}$, two orientations of $\log(u)Z^{(1)}$ and $Z^{(1)}$ itself.
	}
we find that it precisely matches the \textit{full} form of $\ca^\text{one-loop}_{\text{sugra}}(x)$, and not just its discontinuity as reported in~\cite{Alday:2017vkk}. Note that in the matching one has to set the quadratic divergence of $\ca^\text{one-loop}_{\text{sugra}}(x)$ to zero since it is the super-leading term $\ca^{(1,0)}$ which takes the role of the one-loop counter-term. Furthermore, the one-loop ambiguity $\alpha$ does not contribute to the bulk-point limit as its denominator $(x-\xb)^{13}$ is sub-leading compared to $(x-\xb)^{15}$ of $\mathcal{H}^{(2)}$, and we therefore do not get any constraint on $\alpha$ from the bulk-point limit.

The next two non-analytic genus-one contributions are the order $(\ap)^4$ and $(\ap)^6$ terms $\ca^{(1,4)}$, $\ca^{(1,6)}$ given in~\eqref{eq:genus1-terms}. The corresponding AdS correlators $\mathcal{H}^{(2,3)}$ and $\mathcal{H}^{(2,5)}$ have been constructed in~\cite{Drummond:2019hel}, and it turns out that a function with a new type of singularity is present. This weight-three function, $f^{(3)}(x,\xb)$, contains the additional letter $x-\xb$ and is therefore beyond the space of SVHPL's, making the computation of its bulk-point limit more involved. We use the \texttt{Mathematica} package \texttt{PolyLogTools}~\cite{Duhr:2019tlz} to perform the analytic continuations and, introducing the parametrisation $\epsilon=x-\xb$, to then carefully take the limit $\epsilon\to0$. We find that the function $f^{(3)}$ contributes to the bulk-point limit with
\begin{align}\label{eq:f3_bpl}
	f^{(3)}(x,\xb)~\xrightarrow[]{~\text{bpl}~}~ -8\pi^2\big(2\log(x)+2\log(1-x)+i\pi-3\log(\epsilon)\big).
\end{align}
Note the appearance of an extra logarithmic divergence parametrised by $\log(\epsilon)$ due to the presence of the new letter $x-\xb$. 

Now, we can proceed to check the relation~\eqref{eq:bulk-point_matching} between the bulk-point limit of the correlators $\mathcal{H}^{(2,3)}$, $\mathcal{H}^{(2,5)}$,\footnote{
	Besides $f^{(3)}$, also $Z^{(1)}$ and the two orientations of $\log(u)Z^{(1)}$ contribute to the bulk-point limit.
	}
and the non-analytic genus-one contributions $\ca^{(1,4)}$, $\ca^{(1,6)}$. We find perfect agreement upon identifying the extra logarithmic divergence $\log(\epsilon)$ from the function $f^{(3)}$ with the scale-dependent terms $\log(\frac{\ap\,s}{\mu})$ according to
\begin{align}\label{eq:log_epsilon_identification}
	\log(\epsilon)\simeq \frac{1}{2}\big(\log(x)+\log(1-x)-\log(\tfrac{\ap s}{\mu})+i\pi\big),
\end{align}
for a choice of scale $\mu$. Besides providing the expected matching, the above identification can be further justified by checking that both sides of the equation transform equally under all crossing transformations. Indeed, after recasting~\eqref{eq:log_epsilon_identification} into the form $\log(\epsilon^2)=\log(x)+\log(1-x)-\log(\tfrac{\ap s}{\mu})+i\pi$ and recalling $\epsilon=x-\xb$, one can verify that the proposed identification is consistent with crossing.

Lastly, note that the dependence of~\eqref{eq:log_epsilon_identification} on an {\`a} priori arbitrary scale $\mu$ prevents us from fixing any tree-level ambiguities in the one-loop correlators which might contribute to the bulk-point limit. Take for example $\mathcal{H}^{(2,3)}$, which has been fully fixed up to four ambiguities corresponding to the Mellin amplitudes $\{1,\sigma_2,\sigma_3,\sigma_2^2\}$, where $\sigma_n\equiv s^n+t^n+u^n$. In the bulk-point limit, or equivalently in the formulation of the flat-space limit for Mellin amplitudes, only the overall leading powers in the Mellin variables contribute, which in this case is given by the term $\sigma_2^2$. Now, one can check that in the bulk-point limit the coefficient of $f^{(3)}$ is proportional to $s^4(x^2-x+1)^2\propto\sigma_2^2$, and thus a shift in the scale $\mu$ in~\eqref{eq:log_epsilon_identification} simply corresponds to a redefinition of the coefficient of the ambiguity.

\subsubsection*{Constraints for $\mathcal{H}^{(3)}$ from matching at genus-two}
The leading term at genus two is of order $(\ap)^2$ and corresponds to the genus-two correction of the $\partial^4\mathcal{R}^4$ term. In AdS, this contribution has been determined in~\cite{Chester:2020dja} and one can check that in the bulk-point limit it correctly reproduces the analytic contribution $\ca^{(2,2)}$ given in~\eqref{eq:genus2-terms_1}.

Let us skip forward and focus on the $(\ap)^5$ contribution, which is of particular relevance to us as it provides a non-trivial constraint for the two-loop correlator $\mathcal{H}^{(3)}$. As stated in~\eqref{eq:genus2-terms_2}, one expects two distinct contributions to $\ca^{(2,5)}$: an analytic contribution from $\partial^{10}\mathcal{R}^4\vert_{\text{genus-2}}$ of the form $\sigma_2\sigma_3\propto s^5\,x(1-x)(x^2-x+1)$ and a non-analytic contribution due to the two-loop supergravity amplitude $\ca_{\text{sugra}}^{\text{two-loops}}$, which is given by the sum over crossing orientations of the planar and non-planar double-box integral in 10 dimensions, see e.g.~\cite{Bern:1997nh}. Its finite part has been evaluated in~\cite{Bissi:2020woe}, whose result we use to compute the function of angles $\ca_{\text{sugra}}^{\text{two-loops}}(x)$.\footnote{
	We record the explicit expression for $\ca_{\text{sugra}}^{\text{two-loops}}(x)$ in an ancillary file. It is given in terms of HPL's up to weight 4 with rational coefficients.}
Notably, the $1/\epsilon^2$ pole in dimensional regularisation cancels in the sum of the planar and non-planar integral such that the remaining divergence is only of order $1/\epsilon$. This gives rise to a scale-dependent term with a logarithmic contribution of the form $s^5\log(\frac{\ap s}{\mu})$. We find that its coefficient is proportional to $x(1-x)(x^2-x+1)$, which is in agreement with the expected presence of the analytic contribution since a change in the scale $\mu$ simply amounts to a shift in the coefficient of $\partial^{10}\mathcal{R}^4\vert_{\text{genus-2}}$.

Note that the scale-dependent contribution to $\ca^{\text{two-loops}}_{\text{sugra}}$ has a non-trivial consequence for the corresponding AdS correlator. In the bulk-point limit, the two-loop supergravity correlator $\mathcal{H}^{(3)}$ maps to the genus-two $(\ap)^5$ term $\ca^{(2,5)}$, and in order to match the scale-dependent logarithm we need a non-zero contribution from $f^{(3)}(x,\xb)$ to the bulk-point limit of $\mathcal{H}^{(3)}$. We will therefore necessarily need to include the letter $x-\xb$ in our ansatz for $\mathcal{H}^{(3)}$ at least to weight 3, in contrast to the one-loop correlator $\mathcal{H}^{(2)}$ which is given in terms of SVHPL's alone. Furthermore, as in the case for the one-loop string correlators discussed earlier, the dependence of the identification~\eqref{eq:log_epsilon_identification} on an arbitrary scale $\mu$ (as well as our ignorance of the precise coefficient of $\partial^{10}\mathcal{R}^4\vert_{\text{genus-2}}$) will prevent us from fixing the expected tree-level ambiguity $\sigma_2\sigma_3$ in the two-loop correlator $\mathcal{H}^{(3)}$.

\section{Bootstrapping the two-loop correlator $\mathcal{H}^{(3)}$}\label{sec:two_loops}\setcounter{equation}{0}
Before explaining the details of our construction of the two-loop supergravity contribution, it is useful to first step back and review the known results at preceding orders. The structures appearing in the tree-level and one-loop correlators, $\mathcal{H}^{(1)}$ and $\mathcal{H}^{(2)}$, will then guide us to make an educated ansatz for the two-loop correlator $\mathcal{H}^{(3)}$.

\subsection{Review of tree-level and one-loop}
The derivation of the tree-level correlator $\mathcal{H}^{(1)}$ dates back to the early days of the AdS/CFT correspondence and belongs to the first explicit results of an amplitude on AdS. In our conventions, it is given by~\cite{Arutyunov:2000py,Dolan:2001tt}
\begin{align}\label{eq:H1}
	\mathcal{H}^{(1)} = -16u^2\dbar{2422}.
\end{align}
In hindsight, this result can be bootstrapped from the following considerations: the basis function for the top-weight part is given by the first member of zigzag integrals, $Z^{(1)}$. The remaining lower-weight basis elements are then simply given by $\log(u)$, $\log(v)$ and $1$. Each of these four functions is multiplied by a rational coefficient function with denominator power $(x-\xb)^7$. A subtlety at tree-level is that one also has to allow for a (single) power of $v=(1-x)(1-\xb)$ in the denominator in order to ensure cancellation of certain protected twist 2 contributions. Finally, requiring crossing symmetry, no unphysical poles at $x=\xb$ and cancellation of the twist 2 sector fully fixes $\mathcal{H}^{(1)}$ and precisely yields the result given above.

Considerably more involved is the one-loop correlator $\mathcal{H}^{(2)}$. Explicitly given in its full form in position space in~\cite{Aprile:2017bgs}, it has later been recast in a much simpler form by using the fact that its leading log can be written as $\d8$ acting on a simpler object, as explained in Section~\ref{sec:delta8_and_leading-log}. Remarkably, one finds that this property of the leading log extends to the \textit{full} correlator, such that $\mathcal{H}^{(2)}$ can be written as $\d8$ acting on a `preamplitude' $\L^{(2)}$ at the expense of having to add some amount of the tree-level correlator $\mathcal{H}^{(1)}$~\cite{Aprile:2019rep},
\begin{align}\label{eq:H2}
	\mathcal{H}^{(2)} = \frac{1}{u^2} \d8 \L^{(2)} + \mathcal{H}^{(1)}.
\end{align}
Let us emphasise again that this achieves a remarkable simplification: the coefficient functions of the transcendental basis elements in $\mathcal{H}^{(2)}$ come with maximal denominator power $(x-\xb)^{15}$, whereas the corresponding coefficient functions in $\L^{(2)}$ have at most denominator power $(x-\xb)^{7}$, with the extra $8$ powers being supplied through the action of the 8-th order differential operator $\d8$. In that regard, the coefficient functions at one-loop order turn out to be of the same complexity as the tree-level ones.

The basis of transcendental functions appearing in $\L^{(2)}$ is again consistent with our previous observations: at top-weight (weight 4 in this case), it is given by the three independent orientations of the zigzag integral $Z^{(2)}$, completed by all lower-weight functions with no $\log^3(u)$ contribution to any channel. In contrast to the tree-level case however, the bootstrap constraints described above are not sufficient to entirely fix the full one-loop correlator. Instead, one is left with one remaining free parameter, $\alpha$, which comes with a tree-level like function given by $u^2\dbar{4444}$, corresponding to a constant Mellin amplitude. The presence of such a tree-level ambiguity is related to the super-leading counter-term $\mathcal{R}^4\vert_{\text{genus-1}}$, and its value is determined within the full type IIB string theory effective action on AdS$_5\times$S$^5$. However, our bootstrap computation is not able to fix it and one has to resort to other methods. For instance, as shown in~\cite{Chester:2019pvm}, it is possible to determine it by using supersymmetric localisation techniques, yielding the value $\alpha=60$.

Interestingly, it is possible to write the one-loop ambiguity $\alpha$ as part of the preamplitude $\L^{(2)}$, which is a remarkable property of the corresponding $\dbar{}$-function. Note that this term contributes non-analytically with a single spin $\ell=0$ conformal block to the one-loop anomalous dimension $\gamma^{(2)}$, as recorded in equation~\eqref{eq:gamma2} for the twist 4 case. The presence of the $\alpha$-parameter at one-loop order will in turn induce a non-analytic spin 0 contribution to the $\log^2(u)$-part of the two-loop correlator.

\subsection{More on $\d8$: symmetries under crossing}\label{sec:d8_symmetries}
As this will become important in what comes next, let us briefly comment on the crossing properties of the $\d8$ operator, which we introduced earlier in Section~\ref{sec:leading-logs}. For convenience, we repeat its definition:
\begin{align}\label{eq:d8_def2}
	\d8 = \frac{u^4}{(x-\xb)}\partial_{\xb}^2(1-\xb)^2\partial_{\xb}\partial_x^2(1-x)^2\partial_x^2(x-\xb).
\end{align}
Recall that $\d8$ was useful in the context of the leading log of $\mathcal{H}^{(n)}$, since pulling out $n-1$ powers of $\d8$ drastically simplifies the leading log. In order for this to be consistent with the $1\leftrightarrow2$ exchange-symmetry of the OPE decomposition, this operator itself needs to respect that symmetry. Indeed, one can check that under the corresponding crossing transformation $x\mapsto x'=\frac{x}{x-1}$, the $\d8$ operator (and consequently any power of it) is left invariant
\begin{align}
	\big(\d8\big)^k~\xrightarrow[]{~x\,\rightarrow\,x'~}~~\big(\d8\big)^k.
\end{align}
Remarkably, an accidental enhancement of crossing symmetry occurs for the case when $k=2$. We find that
\begin{align}
	\big(\d8\big)^2~\xrightarrow[]{~x\,\rightarrow\,1-x~}~~\frac{u^4}{v^4}\,\big(\d8\big)^2,
\end{align}
while no other power of $\d8$ obeys this extra symmetry. As a simple consequence, the combination $\frac{1}{u^2}(\d8)^2$ has the same crossing symmetries as the correlator $\mathcal{H}(u,v)$ itself, see equation~\eqref{eq:crossing_H}. This enhancement to full crossing symmetry is particularly useful in the construction of the two-loop amplitude $\mathcal{H}^{(3)}$, as it allows the preamplitude (introduced next) to be made fully crossing invariant from the get go. In contrast, the preamplitude of the one-loop correlator obeys only the $x\mapsto x'$ crossing symmetry.

\subsection{A minimal ansatz for the two-loop correlator}
Based on the fact that $(\d8)^2$ can be pulled out from the leading log, and further motivated by the simple structure of the one-loop correlator as given in equation~\eqref{eq:H2}, we start with the following minimal ansatz for the two-loop correlator,\footnote{
	We have found this structure to be the simplest ansatz possible which satisfies all of the imposed bootstrap constraints (to be discussed below), justifying the name `minimal ansatz'. For an exploration of different modifications of the minimal ansatz, see Section~\ref{sec:wider_ansatz}.
	}
\begin{align}\label{eq:H3_ansatz_minimal}
	\mathcal{H}^{(3)} = \frac{1}{u^2}(\d8)^2\,\P^{(3)} + a_2\mathcal{H}^{(2)} + a_1\mathcal{H}^{(1)},
\end{align}
with $\P^{(3)}$ denoting the two-loop preamplitude we would like to compute. Let us point out that the structure of our minimal ansatz is of the same form as proposed in the recent work by Huang and Yuan~\cite{Huang:2021xws}, albeit different in some details. We will comment on the precise differences of the final results in Section~\ref{sec:comparison_with_Huang/Yuan}.

On general grounds, the two-loop preamplitude is of the form
\begin{align}\label{eq:ansatz_p3}
	\P^{(3)}(x,\xb) = \sum_{i}\frac{p_i(x,\xb)}{(x-\xb)^{d_i}}\,\mathcal{Q}_i(x,\xb),\quad\text{with } p_i(x,\xb)=\sum_{n=0}^{d_i}\sum_{m=n}^{d_i} a_{n,m}^{(i)}\left(x^n \xb^m+x^m \xb^n\right),
\end{align}
where the $\mathcal{Q}_i$ are pure transcendental functions (forming the basis $\mathcal{Q}$ specified below) with coefficient polynomials $p_i(x,\xb)$ containing free parameters $a_{n,m}^{(i)}$. The denominator powers are given by $d_i=7$ ($d_i=6$) when the corresponding function $\mathcal{Q}_i(x,\xb)$ is antisymmetric (symmetric) under the exchange symmetry $x\leftrightarrow\xb$. This ensures that overall the preamplitude is a symmetric function of $(x,\xb)$.

Let us now proceed to specify our basis of transcendental functions. Our basis contains SVHPL's built from the alphabet $\{x,\xb,1-x,1-\xb\}$ up to transcendental weight 6, subject to the following conditions:
\begin{itemize}
	\item The functions at weight 6 are constrained by the observations on the leading log discussed in Section~\ref{sec:leading-logs}. At two-loop order, the top-weight part is given by the zizag-integral $Z^{(3)}$ which provides the correct leading-log contribution to $\mathcal{H}^{(3)}$. In addition, we include all other weight 6 functions with no further $\log^3(u)$ contributions in any orientation and which vanish at $x=\xb$.\footnote{
		This is a necessary condition for ensuring the cancellation of spurious poles at $x=\xb$ discussed later.}
	It turns out that these two conditions constrain the additional weight 6 functions to be antisymmetric.
	
	\item For functions of weight $w=0,1,\ldots5$, the only selection criterion we impose is the fact that the OPE predicts the leading logarithmic divergence to be of order $\log^3(u)$. Below weight 6, we will thus include all functions (both symmetric and antisymmetric) with no $\log^4(u)$ contributions to any channel. Recall that in Section~\ref{sec:zigzag} we argued that all functions contributing to the leading log are in principle further constrained to be given by zigzags or derivatives thereof. However, for simplicity we will refrain from imposing this condition here, at the expense of possibly working with a slightly wider transcendental basis than necessary. As a consequence, we will find that any such extra functions have their coefficients set to zero in the final result.
	
	\item Lastly, we need to consider the possibility of including functions with the additional letter $x-\xb$, which goes beyond the space of SVHPL's. As shown in~\cite{Drummond:2019hel}, the one-loop $1/\l$ corrections $\mathcal{H}^{(2,m)}$ necessarily contain such a function at weight 3, denoted by $f^{(3)}(x,\xb)$ therein.\footnote{
		In fact, $f^{(3)}(x,\xb)$ is the unique such new function at weight 3. This is not the case at higher weights any more, however.}
	In particular, its presence is required because the one-loop string corrections have \textit{finite} spin contributions to the leading log, in contrast to the infinite spin support of $\mathcal{H}^{(2)}\vert_{\log^2(u)}$. Now, it is for the same reason that we need to include it as part of our transcendental basis for $\mathcal{H}^{(3)}$: the one-loop ambiguity $\alpha$ from $\mathcal{H}^{(2)}$ induces a one-loop like contribution to the two-loop supergravity correlator, which is exactly of the form of $\mathcal{H}^{(2,3)}$.\footnote{
		Recall there is a similar term at order $a^3\l^{\frac{1}{2}}$, which is a one-loop counter-term coming from a diagram containing a supergravity and a $\mathcal{R}^4\vert_{\text{genus-1}}$ vertex. This term is in fact super-leading with respect to the two-loop supergravity correlator, leading us to anticipate the presence of such a contribution also in $\mathcal{H}^{(3)}$.} 
	We therefore expect $f^{(3)}$ to contribute at two-loop supergravity level, in agreement with arguments from the bulk-point limit.	
	On the other hand, we exclude higher-weight functions with letter $x-\xb$ for now and revisit such a possibility later in Section~\ref{sec:wider_ansatz_w4}.
\end{itemize}
The above considerations leave us with a basis of 73 independent functions $\mathcal{Q}_i(x,\xb)$. A schematic overview of the basis $\mathcal{Q}$ is presented in Table~\ref{tab:basis_overview}, while the precise definitions and symmetry properties of all the functions are spelled out in great detail in Appendix~\ref{app:basis}. Note that we consider explicit $\zeta$-values as independent basis elements, such that the free coefficients $a_{n,m}^{(i)}$ in the preamplitude~\eqref{eq:ansatz_p3} are simply rational numbers. Before having imposed any constraints on $\P^{(3)}$, our initial ansatz for the preamplitude contains 2308 free parameters.

\renewcommand{\arraystretch}{1.3}
\begin{table}
	\begin{center}
		\begin{tabular}{c|c|c|c}
			$w$ & $x\leftrightarrow\xb$ & $\mathcal{Q}_i(x,\xb)$ & total \\\hline
			6	   & $-$ & $6\times Z^{(3)},~A^{(6)},~3\times B^{(6)},~\zeta_3f^{(3)},~2\times\zeta_3\log(u)Z^{(1)}$ & 13 \\
			& $+$ & - & 0 \\\hline
			5	   & $-$ & $6\times\widetilde{\Psi}^{(3)},~3\times\widetilde{\Pi}^{(5)},~\zeta_3Z^{(1)}$ & 10 \\
			& $+$ & $6\times\Psi^{(3)},~6\times\Pi^{(5)},~\Omega^{(5)},~2\times\log(u)(Z^{(1)})^2,~3\times\zeta_3\log^2(u)$ & 18 \\\hline
			4	   & $-$ & $3\times\log^2(u)Z^{(1)},~3\times Z^{(2)}$ & 6 \\
			& $+$ & $6\times\Upsilon^{(3)},~2\times\log^3(u)\log(v),~(Z^{(1)})^2$ & 9 \\\hline
			3	   & $-$ & $f^{(3)},~2\times\log(u)Z^{(1)}$ & 3 \\
			& $+$ & $4\times\log^3(u),~3\times\Psi^{(2)}$ & 7 \\\hline
			2	   & $-$ & $Z^{(1)}$ & 1 \\
			& $+$ & $3\times\log^2(u)$ & 3 \\\hline
			1	   & $-$ & - & 0 \\
			& $+$ & $2\times\log(u)$ & 2 \\\hline
			0	   & $-$ & - & 0 \\
			& $+$ & $1$ & 1
		\end{tabular}
	\end{center}
	\caption{Overview of the basis of transcendental functions $\mathcal{Q}$, ordered by their transcendental weight $w$. At each weight, we classify the basis elements according to their symmetry under $x\leftrightarrow\xb$ exchange, distinguishing antisymmetric ($-$) from symmetric ($+$) functions. The third column explicitly lists the functions $\mathcal{Q}_i(x,\xb)$ together with their number of independent orientations under crossing (e.g. at weight $w=1$, the entry $2\times\log(u)$ means there exist 2 orientations of this basis element, given by $\log(u)$ and $\log(v)$ in this case). Finally, in the last column we give the total number of functions in each category.}
	\label{tab:basis_overview}
\end{table}

\subsection{Bootstrapping $\mathcal{H}^{(3)}$: constraints}\label{sec:constraints}
We now turn to the description of the constraints which the ansatz~\eqref{eq:H3_ansatz_minimal} for $\mathcal{H}^{(3)}$ needs to obey. First of all, there are certain constraints which can be imposed directly on the preamplitude $\P^{(3)}$:
\begin{enumerate}[label={(\arabic*)}]
	\item As a consequence of the simple transformation properties of $(\d8)^2$ under crossing, we can take $\P^{(3)}$ to be fully crossing symmetric:
		\begin{align}\label{eq:crossing_P3}
			\P^{(3)}(x,\xb)=\P^{(3)}(x',\xb')=\P^{(3)}(1-x,1-\xb).
		\end{align}

	\item We can furthermore impose the matching of the predicted leading log directly on the preamplitude $\P^{(3)}$, as discussed in Section~\ref{sec:leading-logs}. This amounts to imposing:
	\begin{align}
	\begin{split}
		\mathcal{P}^{(3)}(x,\bar{x}) = \,\, &\mathcal{G}^{(3)}(x,\bar{x}) + \mathcal{G}^{(3)}(1-x,1-\bar{x}) + \mathcal{G}^{(3)}(1-x',1-\bar{x}')\\
		& + \text{ terms with no $\log^3 u$ in any channel,}
	\end{split}
	\end{align}
	where $\mathcal{G}^{(3)}$ is the combination of zigzags and derivatives introduced in (\ref{cG}). However, in practice we impose this only for the top-weight part for simplicity and match the rest against our ansatz.
	
	\item The full correlator $\mathcal{H}^{(3)}(u,v)$ is expected to be non-singular at $x=\xb$, as such poles would be at unphysical locations. Note that since the (repeated) application of $\d8$ does not create any new poles at $x=\xb$, we can directly impose the cancellation of the 7 explicit poles from the denominators in \eqref{eq:ansatz_p3} within the preamplitude $\P^{(3)}$.
\end{enumerate}
At this stage, the ansatz for $\mathcal{H}^{(3)}$ obeys the correct crossing symmetries, has no poles at $x=\xb$ and by construction matches the correct leading log. However, we are still left with a total of 82 free parameters (80 from $\P^{(3)}$ together with $a_1$ and $a_2$). On the other hand, there are further constraints which $\mathcal{H}^{(3)}$ needs to satisfy:
\begin{enumerate}[label={(\arabic*)}]\setcounter{enumi}{3}
	\item Cancellation of all contributions with twists below 4. As in the one-loop case, the presence of the tree-level expression $\mathcal{H}^{(1)}$ in our ansatz is required in order to satisfy this constraint. This step gives 60 constraints on the free parameters, and in particular we find that $a_1$ is fixed to take the value $a_1=-1$.

	\item Matching the predicted $\log^2(u)$ contribution at twist 4. According to the OPE expansion~\eqref{eq:logu_stratification}, this term is given in terms of one-loop and tree-level OPE data by the combination
	\begin{align}\label{eq:log2u_twist4_sum}
		x^2g(\xb)\equiv\sum_{\ell}\left(\frac{1}{2}\,A_{2,\ell}^{(1)}\,\big(\gamma_{2,\ell}^{(1)}\big)^2+A_{2,\ell}^{(0)}\,\gamma_{2,\ell}^{(1)}\,\gamma_{2,\ell}^{(2)}+A_{2,\ell}^{(0)}\big(\gamma_{2,\ell}^{(1)}\big)^3 \partial_{\Delta}\right)G_{2,\ell}(x,\xb)\,,
	\end{align}
	where we have kept only twist 4 contributions. Performing the sum over even spins~$\ell$ we obtain the explicit one-variable function $g(\xb)$ recorded in Appendix~\ref{app:log^2u_twist4}. Notably, it contains a non-analytic spin 0 contribution from the one-loop ambiguity $\alpha$. We find 8 further constraints from this matching, one of which determines $a_2=5$.

	\item Matching the flat-space contribution $\ca^{(2,5)}$. As explained in Section~\ref{sec:bulk_point}, the correlator $\mathcal{H}^{(3)}$ is mapped to the two-loop supergravity amplitude in the bulk-point limit. We fix 6 more free parameters by matching against the explicit result for $\ca_{\text{sugra}}^{\text{two-loops}}(x)$ from reference~\cite{Bissi:2020woe}, whilst keeping the ambiguities corresponding to the $\partial^{10}\mathcal{R}^4\vert_{\text{genus-2}}$ correction term.\footnote{
		As already emphasised in Section~\ref{sec:bulk_point}, we perform this matching at the level of the \textit{full} amplitude, not only its discontinuity. While this does not result in giving more constraints, being able to match a complicated expression involving HPL's of up to weight 4 nevertheless constitutes a highly non-trivial consistency check between the result for the flat-space amplitude from~\cite{Bissi:2020woe} and our construction of the CFT correlator.}
\end{enumerate}

\subsection{Bootstrapping $\mathcal{H}^{(3)}$: results}\label{sec:results}
After having imposed the constraints (1) -- (6) described above, we find that the two-loop correlator $\mathcal{H}^{(3)}$ is completely fixed up to only 8 remaining free parameters, which all correspond to tree-level ambiguities. Our final result takes the form
\begin{align}\label{eq:H3_result}
	\mathcal{H}^{(3)} = \frac{1}{u^2}(\d8)^2\,\P^{(3)} + 5\,\mathcal{H}^{(2)} - \mathcal{H}^{(1)},
\end{align}
with the tree-level and one-loop correlators given in~\eqref{eq:H1} and~\eqref{eq:H2}, respectively. Due to the complexity of the coefficient functions $p_i(x,\xb)$, we attach our result for the preamplitude $\mathcal{P}^{(3)}$ and also the full correlator $\mathcal{H}^{(3)}$ (for some choice of ambiguities) in an ancillary file to the \texttt{arXiv} submission. We find it remarkable that our highly constrained minimal ansatz is able to satisfy all of the imposed constraints, and we emphasise that the only free parameters left are of the form of tree-level contact diagrams.

We observe that not all of the transcendental functions $\mathcal{Q}_i(x,\xb)$ which were included in the initial ansatz actually contribute to the final result. In the notation of Table~\ref{tab:basis_overview} and organised by their transcendental weight $w$, the 10 basis elements with vanishing coefficients (both in $\P^{(3)}$ and $\mathcal{H}^{(3)})$ read
\begin{itemize}
	\item $w=6$: $\quad3\times B^{(6)},~2\times\zeta_3\log(u)Z^{(1)}$,
	\item $w=5$: $\quad\Omega^{(5)},~2\times\log(u)(Z^{(1)})^2$,
	\item $w=4$: $\quad2\times\log^3(u)\log(v)$.
\end{itemize}
Note that the functions of weight 4 and 5 listed above all have $\log^3(u)$ contributions and the vanishing of their coefficients is a consequence of the observation made in Section~\ref{sec:zigzag} that the leading log is fully captured by the zigzag functions $Z^{(n)}$ and derivatives thereof. Indeed, one can check that the functions $\Omega^{(5)}$, $\log(u)(Z^{(1)})^2$ and $\log^3(u)\log(v)$ can not be expressed in terms of derivatives of zigzags, and hence they do not appear in the two-loop correlator. On the other hand, the same argument does not apply to the weight 6 functions listed above as they do not contribute to the leading log. Currently, we do not have an explanation for the unexpected vanishing of these particular functions and it would be very interesting to understand the principle behind this observation.

Let us now turn to remaining tree-level ambiguities in $\mathcal{H}^{(3)}$. Schematically, the 8 free parameters split into $4\times\dbar{}$ and $4\times\zeta_3\dbar{}\,$,\footnote{%
	Recall that we do not allow for explicit $\zeta$-values in the coefficient polynomials. Instead, all $\zeta$-values are treated as separate basis elements of $\mathcal{Q}$, explaining why $\dbar{}$ and $\zeta_3\dbar{}$ appear as independent functions.}
whose coefficients we parametrise by $d_j$ and $e_j$, with $j=1,2,3,4$. Such tree-level ambiguities are best described in terms of their Mellin amplitudes, and we find they are given by the following linear combinations of monomials $\{1,\sigma_2,\sigma_3,\sigma_2^2,\sigma_2\sigma_3\}$,
\begin{align}\label{eq:ambiguities_H3}
	d_1 \simeq \sigma_2-\tfrac{16}{7},\qquad	d_2 \simeq \sigma_3+\tfrac{32}{7},\qquad d_3 \simeq \sigma_2^2-\tfrac{128}{7},\qquad d_4 \simeq \sigma_2\sigma_3-\tfrac{256}{7},
\end{align}
and the other 4 are simply related by $e_j=\zeta_3\,d_j$. In fact, these are precisely the expected ambiguities at this order as they are the ones which contribute to the $\partial^{10}\mathcal{R}^4\vert_{\text{genus-2}}$ term in AdS. In the conformal block expansion, they contribute to the $\log(u)$-part with finite spin support for spins $\ell=0,2,4$ only, and hence there are no ambiguities with infinite spin support in our two-loop result~\eqref{eq:H3_result}.

Interestingly, as in the one-loop case, these unfixed parameters are all written as part of the preamplitude $\P^{(3)}$, which is a non-trivial statement about the corresponding $\dbar{}$-functions. However, note that the Mellin monomial `$1$' does not appear as an independent ambiguity in~\eqref{eq:ambiguities_H3} since it can not be written as $(\d8)^2$ acting on a preamplitude.\footnote{
	This can be easily seen in the positions space representation: the $\dbar{}$-function corresponding to the Mellin amplitude `$1$' is $u^2\dbar{4444}$ which has denominator power $(x-\xb)^{13}$. This is too small as the minimal denominator power generated by applying $(\d8)^2$ is $(x-\xb)^{17}$.
	}
In light of the vanishing of the $\mathcal{R}^4\vert_{\text{genus-2}}$ contribution (both in AdS$_5\times$S$^5$ and in the flat-space string amplitude), there is no super-leading two-loop $\mathcal{R}^4$ counterterm. This is consistent with $u^2\dbar{4444}$ not being an independent ambiguity of the two-loop correlator $\mathcal{H}^{(3)}$ and we speculate that~\eqref{eq:ambiguities_H3} indeed gives to most general form of allowed ambiguities.

Lastly, let us mention how the one-loop ambiguity $\alpha$ enters our result for $\mathcal{H}^{(3)}$. To this end, we have left it as a free coefficient in our final expression even though its value is actually fixed.\footnote{
	Recall that the localisation computation of~\cite{Chester:2019pvm} yields $\alpha=60$.
	}
As anticipated, we find that $\alpha$ is simply proportional to the one-loop string correction $\mathcal{H}^{(2,3)}$ (with its tree-level ambiguities fixed to some values). Furthermore, just like the ambiguities~\eqref{eq:ambiguities_H3}, it turns out that $\mathcal{H}^{(2,3)}$ can in fact be written as part of the preamplitude, i.e. $\mathcal{H}^{(2,3)}=\frac{1}{u^2}(\d8)^2\,\P^{(2,3)}$ for some $\P^{(2,3)}$. This is surprising as $\mathcal{H}^{(2,3)}$ is really a one-loop amplitude and naively one would expect that only a single power of $\d8$ could be pulled out.

\subsection{Comparison with the results of Huang and Yuan}\label{sec:comparison_with_Huang/Yuan}
Let us denote the result obtained in~\cite{Huang:2021xws} by $\mathcal{H}^{(3)}_{\text{HY}}$.\footnote{
	Here $\mathcal{H}^{(3)}_{\text{HY}}$ refers to the main part of their amplitude which we take from their ancillary file.}
Our result for $\mathcal{H}^{(3)}$ agrees with the main part of their correlator upon setting their free parameter $\mathcal{X}=0$. In particular, after changing to our conventions, we find that both results agree on the precise coefficients of the $\mathcal{H}^{(1)}$ and $\mathcal{H}^{(2)}$ contributions to $\mathcal{H}^{(3)}$. The exact difference between our two results reads
\begin{align}\label{eq:comparison_HY}
	\mathcal{H}^{(3)}_{\text{our}} - \mathcal{H}^{(3)}_{\text{HY}}\vert_{\mathcal{X}=0} = \tfrac{36}{7}\,\zeta_3\mathcal{H}^{(2,3)} - \big(\tfrac{5849}{1008}-\tfrac{\alpha }{240}\big)\,\mathcal{H}^{(2,3)} + \big(\dbar{}\text{- and }\zeta_3\dbar{}\text{-ambiguities}\big),
\end{align}
where the last term stands schematically for the tree-level ambiguities expected from the $\partial^{10}\mathcal{R}^4\vert_{\text{genus-2}}$ correction, i.e. terms with Mellin amplitudes  $\{1,\sigma_2,\sigma_3,\sigma_2^2,\sigma_2\sigma_3\}$. The above difference in our respective results is consistent with equation (37) of~\cite{Huang:2021xws}, since (apart from the usual tree-level ambiguities mentioned before) additional one-loop ambiguities have been left unfixed.

The condition that the free parameter $\mathcal{X}$ in their correlator $\mathcal{H}^{(3)}_{\text{HY}}$ is required to vanish in the comparison~\eqref{eq:comparison_HY} is explained when considering the corresponding preamplitude $\P^{(3)}_{HY}$, which we have reconstructed from their expression for the full correlator. We find that the contribution of $\mathcal{X}$ is sourced by the presence of new weight 4 functions containing the letter $x-\xb$ in the preamplitude $\P^{(3)}_{HY}$, which are however annihilated once $\d8$ is applied such that their contributions to the full amplitude $\mathcal{H}^{(3)}_{HY}$ vanish. Since we have not included such new functions at weight 4 in our preamplitude, we necessarily find that $\mathcal{X}=0$ and we thereby fix their free parameter.

Note that the vanishing of $\mathcal{X}$ is conjectured in~\cite{Huang:2021xws}, but for a different reason: the authors state that the condition $\mathcal{X}=0$ follows from imposing full crossing invariance of the preamplitude $\P^{(3)}_{HY}$, which at first sight seems to be inconsistent with the enhancement of crossing symmetry for the particular case of $(\d8)^2$, recall the discussion in Section~\ref{sec:d8_symmetries}. Indeed, we find that $\P^{(3)}_{HY}$ can be made fully crossing symmetric regardless of the value of $\mathcal{X}$ by adding terms which are in the kernel of $(\d8)^2$. However, these extra terms come with $\log^3(u)$ contributions which would spoil the imposed matching of the leading log at the level of the preamplitude. In other words, the presence of the free parameter $\mathcal{X}$ is excluded by imposing full crossing symmetry \textit{together} with matching of the leading log on $\P^{(3)}$, i.e. constraints (1) and (2) of Section~\ref{sec:constraints}. Of course, this relies on insisting on a `canonical' form for the leading-log preamplitude as defined by the OPE resummation or, equivalently, the ten-dimensional conformal symmetry which diagonalises the tree-level mixing problem.

In summary, as shown in equation~\eqref{eq:comparison_HY}, our correlator $\mathcal{H}^{(3)}$ is in agreement with the results of~\cite{Huang:2021xws}, up to the following details:
\begin{itemize}
	\item Due to the absence of weight 4 functions with letter $x-\xb$ in our minimal ansatz, we determine their free parameter $\mathcal{X}=0$.
	
	\item We fix the one-loop ambiguities present in equation (37) of~\cite{Huang:2021xws} by carefully tracking the contribution of the one-loop ambiguity $\alpha$ to order $a^3$. This explains the presence of the $\mathcal{H}^{(2,3)}$ terms in the comparison \eqref{eq:comparison_HY}.
	
	\item Some shifts in the tree-level ambiguities are required in the comparison, hence the presence of the $\dbar{}$ terms in~\eqref{eq:comparison_HY}. However, note that since we write the ambiguities as part of the preamplitude, we no longer consider $u^2\dbar{4444}$ as an independent ambiguity in $\mathcal{H}^{(3)}$.
\end{itemize}

\subsection{Comments on triple-trace contributions to $\mathcal{H}^{(3)}$}\label{sec:triple-trace}
At some point in the perturbative large $N$ expansion new operators besides the usual tower of double-trace operators $\cO_{t,\ell,i}$ will contribute. It is expected that the next relevant family of operators is given by triple-trace operators, and arguments for their presence at two-loop order have been given by considering unitarity cuts of the corresponding flat-space amplitudes~\cite{Bissi:2020wtv,Bissi:2020woe} or also directly at the level of AdS diagrams, see, e.g.~\cite{Meltzer:2019nbs}.

Here, we present an argument based on consistency of the OPE decomposition. In particular, we will show that there are contributions from new operators in the $su(4)$ singlet representation to the $\log^2(u)$ term of $\mathcal{H}^{(3)}$, starting from twist 6. The argument goes as follows:

Let us consider the twist 6 $\log^2(u)$ contribution at two-loop order, which according to the expansion~\eqref{eq:logu_stratification} is entirely determined in terms of one-loop and tree-level OPE data. Now, \textit{assuming} that only double-trace operators contribute\footnote{
	Twist 6 corresponds to $t=3$ and there are two degenerate double-trace operators $\cO_{3,\ell,i}$ with $i=1,2$, schematically given by (linear combinations of) $\cO_2\square\partial^{\ell}\cO_2$ and $\cO_3\partial^{\ell}\cO_3$, see the discussion above~\eqref{eq:double_trace_operators}.}
we proceed to resolve the order $a^2$ mixing problem, for which besides $\mathcal{H}^{(2)}$ we also need to use information from the one-loop $\langle\cO_2\cO_2\cO_3\cO_3\rangle$ and $\langle\cO_3\cO_3\cO_3\cO_3\rangle$ correlators determined in~\cite{Aprile:2017qoy,Aprile:2019rep}. The solution of this one-loop mixing problem yields explicit expressions for $\gamma_{3,\ell,i}^{(2)}$ and $A^{(1)}_{3,\ell,i}$ for $i=1,2$, which we then use to compute the double-trace prediction for the twist 6 $\log^2(u)$ part at order $a^3$. However, one finds that this does \textit{not} match the actual result extracted from our result for $\mathcal{H}^{(3)}$ and we are led to conclude that more operators besides the two double-trace operators enter the one-loop mixing problem at twist 6.

Based on intuition from large $N$ counting, we presume these additional operators are given by triple-trace operator which will necessarily mix with the double-trace ones.\footnote{
	Mixing between operators with different trace-structures has already been observed in the case of so-called single-particle operators $\mathcal{O}_p$~\cite{Aprile:2018efk,Aprile:2020uxk}, which are defined as the $\mathcal{N}=4$ half-BPS operators dual to single-particle states of AdS$_5\times$S$^5$ supergravity. The first case where half-BPS single-, double- and triple-trace operators mix is $\cO_6$, which in the $SU(N)$ theory is of the form (see equation (15) of~\cite{Aprile:2020uxk})
	\begin{align}
		\cO_6 = T_6+ \beta_1\,T_{3,3} + \beta_2\,T_{4,2} + \beta_3\,T_{2,2,2},
	\end{align}
	where $T_{p_1,p_2,\ldots,p_n}(x)\equiv T_{p_1}(x)\,T_{p_2}(x)\cdots\,T_{p_n}(x)$, with $T_p(x)\equiv\text{Tr}(\Phi(x)^p)$ being the usual dimension $p$ single-trace operator. The coefficients $\beta_i$ are functions of $N$ with large $N$ behaviours $\beta_{1,2}\sim\frac{1}{N}$ and $\beta_3\sim\frac{1}{N^2}$, leading to the pattern that terms with $m$ traces are $1/N$ suppressed with respect to $(m-1)$-trace terms.

	In analogy, we expect the triple-trace admixtures to the exchanged, unprotected double-trace operators to be $1/N$ suppressed.
}
As a consequence, information about their leading-order anomalous dimension is already contained within the four-point correlators at one-loop order $a^2$. 

Lastly, let us mention that the mismatch between the double-trace prediction for the twist 6 $\log^2(u)$ part and the actual expression extracted from $\mathcal{H}^{(3)}$ is already visible in the structure of the contributing transcendental functions: the twist 6 contribution to $\mathcal{H}^{(3)}\vert_{\log^2(u)}$ contains terms with up to $\log^3(v)$ divergences, whereas the twist 4 contribution due to only double-trace exchanges is found to have at most $\log^2(v)$ terms, which can be verified using the expression given in~\eqref{eq:g(xb)}. This is consistent with the observation that sums over the double-trace spectrum typically lead to at most $\log^2(v)$ divergences (for example the leading log to all loop orders, see also the discussion in~\cite{Bissi:2020woe}). We thus interpret the observed $\log^3(v)$ divergences in the two-loop $\log^2(u)$ part at twist 6 as a sign of triple-trace contributions.

\section{Extracting the two-loop anomalous dimension}\label{sec:gamma3}\setcounter{equation}{0}
With the explicit result for $\mathcal{H}^{(3)}$ at hand, we can proceed to extract new CFT data from it. In particular, the quantity of physical interest is the correction to the dimension of the twist 4 double-trace operators $\cO_{2,\ell}\sim\cO_2\partial^{\ell}\cO_2$. Recall that at higher twists mixing between exchanged operators occurs, with twist 4 being the only case where there exists only one unique operator for each spin. This fact allows us to unambiguously determine its two-loop anomalous dimension $\gamma^{(3)}$, which is encoded in the $\log(u)$-part of the two-loop correlator.

In practice, extracting the two-loop anomalous dimension from $\mathcal{H}^{(3)}\vert_{\log(u)}$ for low, finite spins $\ell$ is straightforward: after subtracting the derivative terms specified by the OPE expansion, see equation~\eqref{eq:logu_stratification}, and performing the block decomposition up to some finite cut-off $\ell_{\max}$ (keeping twist 4 contributions only), one ends up with the particular combination of CFT data $(A^{(2)}_{2,\ell}\gamma^{(1)}_{2,\ell}+A^{(1)}_{2,\ell}\gamma^{(2)}_{2,\ell}+A^{(0)}_{2,\ell}\gamma^{(3)}_{2,\ell})$ for spins $\ell=0,2,\ldots,\ell_{\max}$. Since all quantities except $\gamma^{(3)}$ in that combination are known from lower-order calculations, one can thus solve for $\gamma^{(3)}$ spin by spin.

However, obtaining a closed-form expression directly from the above finite spin data is more difficult. In order to find $\gamma^{(3)}$ as an analytic function of spin, we employ the Lorentzian inversion formula of~\cite{Caron-Huot:2017vep}, whose input is the so-called double-discontinuity $\text{dDisc}(\mathcal{H}^{(3)})$. In this formalism, the exact same combination of CFT data as given above is encoded in the double-poles of the inversion integral. Computing these integrals\footnote{
	We made use of the method described in Appendix C of~\cite{Alday:2017vkk} to recursively simplify the integrals. See also Appendix A of reference~\cite{Alday:2017zzv} for a useful list of results for some inversion integrals.}
and solving for the two-loop anomalous dimension, we obtain $\gamma^{(3)}_{2,\ell}$ as an analytic function of $\ell$, which agrees with the finite spin data obtained from the direct OPE expansion described above. For spins $\ell\geq6$, the two-loop anomalous dimension takes the form
\begin{align}\label{eq:gamma3}
	\gamma^{(3)}_{2,\ell} &= c_3\,\big(S_{-3}-S_{3}-2S_{1,-2}+3\zeta_3\big)+c_2\,S_{-2}+c_1\,S_1+c_0+c_{0}^{(a)}+\alpha\,\tilde{\gamma}^{(2,3)}_{2,\ell},
\end{align}
where the $c_i$ are rational functions of spin and $S_{\vec{a}}\equiv S_{\vec{a}}(\ell+3)$ are nested harmonic sums. For integer arguments and weight-vectors $\vec{a}$ containing non-zero integer entries, they are defined recursively by
\begin{align}\label{eq:harmonic_sums}
	S_{a_1,a_2,\ldots,a_n}(m) = \sum_{k=1}^{m} \frac{(\text{sgn}(a_1))^{k}}{k^{\vert a_1\vert}}S_{a_2,\ldots,a_n}(k), \qquad S_{\emptyset}(m)=1.
\end{align}
Note that the coefficient functions $c_i$ have definite sign under the symmetry $\ell\rightarrow-\ell-7$: apart from $c_0^{(a)}$ which is antisymmetric, all other $c_i$ are symmetric and can therefore be expressed in terms of even powers of the Casimir eigenvalue $J^2$, which at twist 4 is given by $J^2=(\ell+3)(\ell+4)$. We will elaborate further on this so-called reciprocity principle later on. The symmetric coefficient functions read
\begin{small}
\begin{align}\label{eq:c_i}
\begin{split}
	c_3 &= \frac{-221184J^2(J^2-2)(J^8-50 J^6-653592 J^4+30292416 J^2+15169835520)}{5 (J^2-6)^2(J^2-12)(J^2-20)(J^2-30)(J^2-42)(J^2-56)(J^2-72)},\\[3pt]
	c_2 &= \frac{-18432\,q_2(J^2)}{(J^2-6)^2(J^2-12)^2(J^2-20)(J^2-30)^2 (J^2-42)(J^2-56)^2 (J^2-72)(J^2-90)(J^2-132)},\\[3pt]
	c_1 &=\frac{-27648J^2 (J^2-2)(J^8+525 J^6+1730258 J^4-79817784 J^2-39925126080 )}{(J^2-6)^2 (J^2-12)(J^2-20)(J^2-30)(J^2-42)(J^2-56)(J^2-72)},\\[3pt]
	c_0 &= \frac{384\,q_0(J^2)}{5J^2(J^2-6)^5(J^2-12)^2(J^2-20)^2(J^2-30)^2(J^2-42)(J^2-56)^2(J^2-72)(J^2-90)(J^2-132)},
\end{split}
\end{align}
\end{small}
where $q_2(J^2)$ and $q_0(J^2)$ are non-factorisable polynomials given by
\begin{align}
\begin{split}
	q_2(J^2) &= J^2\,\big(23 J^{20}-3252 J^{18}-11511408 J^{16}+3632264384 J^{14}\\
			 &~\qquad-187756129296 J^{12}-46028945140416 J^{10}+6505487171987328 J^8\\
			 &~\qquad-303456834615886848 J^6+5448093169711196160 J^4\\
			 &~\qquad-28725812248908349440 J^2+64442560728268800\big),
\end{split}
\end{align}
and
\begin{align}
\begin{split}
	q_0(J^2) &= \big(3363 J^{32}-1629859 J^{30}+1808457782 J^{28}-530500662732 J^{26}\\
			 &\quad+34917254916536 J^{24}+5640280310229488 J^{22}-1020874675751115744 J^{20}\\
			 &\quad+64205146187309426112 J^{18}-1998754919048666890368 J^{16}\\
			 &\quad+32877932476802852450304 J^{14}-274537156441056453513216 J^{12}\\
			 &\quad+863756354962716443394048 J^{10}+2400097076662032032956416 J^8\\
			 &\quad-23573917734820546679930880 J^6+46280241034580622311424000 J^4\\
			 &\quad-4409814873370414546944000 J^2-66814046963069091840000\big).
\end{split}
\end{align}
The contribution from the one-loop regulator $\alpha$ is, as expected, proportional to the anomalous dimension induced by the one-loop string correction $\mathcal{H}^{(2,3)}$ . We find $\tilde{\gamma}^{(2,3)}_{2,\ell} = \frac{1}{480\zeta_3}\gamma^{(2,3)}_{2,\ell}$, with $\gamma^{(2,3)}_{2,\ell}$ given by~\cite{Binder:2019jwn},
\begin{align}
\gamma^{(2,3)}_{2,\ell} = -\frac{1658880\zeta_3\, J^2(J^2-2)(J^2+4)(J^2+42)}{(J^2-6)(J^2-12)(J^2-20)(J^2-30)(J^2-42)(J^2-56)},\quad\text{for }\ell\geq6.
\end{align}
Lastly, the antisymmetric contribution $c_0^{(a)}$ to the two-loop anomalous dimension takes the simple form
\begin{align}\label{eq:c0_antisymm}
	c_0^{(a)} = \frac{193536(2\ell+7)(J^4-160 J^2+2380)}{(J^2-6)^4(J^2-20)^2}.
\end{align}
In the remainder of this section, we comment on some interesting features of $\gamma^{(3)}$: we first point out some subtleties of our formula~\eqref{eq:gamma3} at low values of the spin, before discussing the reciprocity principle and the large spin limit.

\subsection{Comments on analyticity in spin of $\gamma^{(3)}$}
While the two-loop anomalous dimension is manifestly finite for spins $\ell\geq10$, there are explicit poles for some low values of the spin $\ell$. In particular, the coefficient functions $c_2$ and $c_0$ contain the factors $(J^2-90)$ and $(J^2-132)$ in their denominators, giving rise to (apparent) poles for spins $\ell=6$ and $\ell=8$, respectively. However, it turns out that the numerator of the combination $c_2\,S_{-2}+c_0$ has a zero at exactly those locations, thus cancelling the apparent divergences! Let us emphasise that this cancellation is highly non-trivial, and it depends on the precise spin dependence of the coefficients $c_2$ and $c_0$. Note that there are no such spurious poles at higher values of the spin and the formula~\eqref{eq:gamma3} is manifestly finite for $\ell\geq10$.

One can evaluate $\gamma^{(3)}$ for spins $\ell=6,8$ by using an analytic continuation of the alternating harmonic sum $S_{-2}(\ell+3)$,\footnote{
	Taking the analytic continuation of $S_{-2}(m)$ from odd arguments $m$, see e.g reference~\cite{Albino:2009ci}, one has
	\begin{align}
		S_{-2}(z)= \frac{1}{4} \Big(\psi ^{(1)}\big(\tfrac{z+2}{2}\big)-\psi ^{(1)}\big(\tfrac{z+1}{2}\big)\Big)-\frac{\zeta_2}{2}.
	\end{align}}
yielding the finite values
\begin{align}
\begin{split}
\gamma^{(3)}_{2,\ell=6} &= -\frac{775569722924539}{79394167125}+\frac{9368130816 \zeta (3)}{1028755}-\frac{409464 \alpha }{54145},\\[3pt]
\gamma^{(3)}_{2,\ell=8} &= \frac{147901898176964147}{170666923100625}-\frac{1161059328 \zeta (3)}{1820105}-\frac{16588 \alpha }{13965}.
\end{split}
\end{align}
We find that these values, obtained through an analytic continuation of harmonic sums, are in perfect agreement with the explicit low spin data which we extracted from the OPE expansion of the $\log(u)$ part of $\mathcal{H}^{(3)}$. 

Another interesting feature of the formula for $\gamma^{(3)}$ is the presence of the factor $(J^2-72)$ in the denominator of all 4 coefficient functions $c_i$, see equation~\eqref{eq:c_i}. This factor provides a pole at $\ell=5$ which is not cancelled any more. Even though this pole is at an unphysical value of the spin (only operators with even spins are exchanged in the OPE), it prevents us from analytically continuing below spin 6.\footnote{
	Exactly the same phenomenon occurs in the one-loop anomalous dimension given in equation~\eqref{eq:gamma2}: $\gamma^{(2)}$ has a pole at $\ell=1$, signalling the presence of a non-analytic spin 0 contribution from the one-loop ambiguity $\alpha$.}
This is consistent with the presence of the tree-level ambiguities $d_j,e_j$ given in equation~\eqref{eq:ambiguities_H3}, which are left unfixed by our bootstrap program: being related to the tree-level $\partial^{10}\mathcal{R}^4$ correction, they contribute non-analytically to the two-loop anomalous dimension precisely for finite spin values $\ell=0,2,4$, and $\gamma^{(3)}$ is expected to be analytic only for $\ell\geq6$.

\subsection{$\gamma^{(3)}$ and the reciprocity principle}
For conformal field theories, the reciprocity principle is the general statement that anomalous dimensions of operators are functions of the conformal spin~\cite{Basso:2006nk}. This is equivalent to the statement that the large spin expansion of operator dimensions contains only \textit{even} inverse powers of the bare Casimir $J^2$~\cite{Alday:2015eya}. After taking into account some shifts due to $\mathcal{N}=4$ supersymmetry, the relevant expression for the Casimir eigenvalue takes the form $J^2=(t+\ell+1)(t+\ell+2)$. When the large spin expansion can be resummed, which is the case in our setting (up to some finite spin ambiguities), this implies a discrete $\mathbb{Z}_2$ symmetry of the operator spectrum under the map $\ell\mapsto-\ell-\tau(\ell,a)-3$, with $\tau$ being the full anomalous twist of the relevant operator.\footnote{
	As pointed out in~\cite{Aprile:2017qoy}, this symmetry should be really thought of as a symmetry of the full operator spectrum rather than individual operators, as in the generic case the shift symmetry transforms many objects non-trivially. For example, the tree-level anomalous dimensions $\gamma^{(1)}_{t,\ell,i}$ of the singlet channel double-trace operators, recorded in equation~\eqref{eq:gamma1}, are mapped into each other according to
	\begin{align}
		\gamma^{(1)}_{t,\ell,i} \mapsto \gamma^{(1)}_{t,-\ell-2t-3,i}=\gamma^{(1)}_{t,\ell,t-i}.
	\end{align}
	In this case, the shift symmetry maps data from the family of operators with degeneracy label $i$ onto the family with label $i'=t-i$.
	}

At twist 4, there is a unique operator for each spin, and the above transformation should leave the dimension $\Delta_{2,\ell}$ of the double-trace operators $\cO_{2,\ell}\sim\cO_2\partial^{\ell}\cO_2$ invariant. Plugging-in the large $N$ expansion of their full anomalous twist $\tau(\ell,a)=2(2+a\gamma^{(1)}+a^2\gamma^{(2)}+\ldots)$ and expanding to order $a^3$, one finds the relations  
\begin{align}\label{eq:gamma_reciprocity}
\begin{split}
	\tfrac{1}{2}\big(\gamma^{(1)}_{2,\ell}-\gamma^{(1)}_{2,-\ell-7}\big) &= 0\,,\\[3pt]
	\tfrac{1}{2}\big(\gamma^{(2)}_{2,\ell}-\gamma^{(2)}_{2,-\ell-7}\big) &= \gamma^{(1)}_{2,\ell}\partial_{\ell}\gamma^{(1)}_{2,-\ell-7}\,,\\[3pt]
	\tfrac{1}{2}\big(\gamma^{(3)}_{2,\ell}-\gamma^{(3)}_{2,-\ell-7}\big) &= \gamma^{(1)}_{2,\ell}\,\partial_{\ell}\gamma^{(2)}_{2,-\ell-7} + \gamma^{(2)}_{2,\ell}\,\partial_{\ell}\gamma^{(1)}_{2,-\ell-7} + (\gamma^{(1)}_{2,\ell})^2\,\partial_{\ell}^2\gamma^{(1)}_{2,-\ell-7}\,.
\end{split}
\end{align}
In words, the antisymmetric contribution to the anomalous dimension at a given order is entirely determined by lower-order data. One can easily verify that the first two lines are satisfied by the tree-level and one-loop anomalous dimensions $\gamma^{(1)}$ and $\gamma^{(2)}$, respectively.

Remarkably, our result for the two-loop anomalous dimension is in perfect agreement with the prediction from the reciprocity symmetry: the only antisymmetric contribution to $\gamma^{(3)}$ is given by $c_0^{(a)}=\tfrac{1}{2}\big(\gamma^{(3)}_{2,\ell}-\gamma^{(3)}_{2,-\ell-7}\big)$ as recorded in equation~\eqref{eq:c0_antisymm}, and indeed we find that $c_0^{(a)}$ satisfies the last line of~\eqref{eq:gamma_reciprocity}.

After subtracting the antisymmetric contribution $c_0^{(a)}$, the large spin expansion of the two-loop anomalous dimension is of the form expected from reciprocity symmetry~\cite{Alday:2015eya},
\begin{align}
	\gamma_{2,\ell}^{(3)}-c_0^{(a)} \approx \frac{\beta_2(\log(J))}{J^4}+\frac{\beta_3(\log(J))}{J^6}+\frac{\beta_4(\log(J))}{J^8}+\ldots,
\end{align}
with only even inverse powers of $J$ appearing in the expansion. The leading coefficient $\beta_2$ is given by
\begin{align}
	\beta_2(\log(J))=\frac{384}{5}\big(3363-45\alpha+460\pi^2-864\zeta_3-24(15+4\pi^2)(\log(J)+\gamma_{E})\big),
\end{align}
with $\gamma_{E}$ being the Euler-Mascheroni constant.

\section{An exploration of wider ans{\"a}tze}\label{sec:wider_ansatz}\setcounter{equation}{0}
There are several different possibilities of extending the minimal ansatz~\eqref{eq:H3_ansatz_minimal}, while still preserving the general structure with powers of $\d8$ acting on simpler preamplitudes which we think is well motivated by properties of the leading log. In this section, we consider two natural extensions of the minimal ansatz:
\begin{enumerate}
	\item We address the possibility of including new weight~4 functions with letter $x-\xb$ in the preamplitude $\P^{(3)}$. Previously, we argued that the presence of such a function at weight 3 is required in order to account for the spin 0 non-analyticity of the $\log^2(u)$ twist 4 prediction, and indeed we find the function $f^{(3)}$ to be an essential part of our result for $\mathcal{H}^{(3)}$. Here, we explore the additional degrees of freedom due to the presence of such new functions at weight 4, which {\`a} priori are not necessary to match our bootstrap constraints.
	
	\item The minimal ansatz~\eqref{eq:H3_ansatz_minimal} is of particular simplicity in the sense that the one-loop correction term is exactly given by the one-loop result $\mathcal{H}^{(2)}$ itself. An alternative would be to generalise this term to $\d8$ acting on a more general one-loop like preamplitude different from $\L^{(2)}$ in~\eqref{eq:H2}. We explore such an extension in Section~\ref{sec:wider_ansatz_P2}.
\end{enumerate}
As we describe in the following, these modifications of the minimal ansatz lead to more free parameters in the final answer, which are no longer of the form of tree-level ambiguities only. In general, such additional free parameters will contribute to the two-loop anomalous dimension $\gamma^{(3)}$ with infinite spin support, although at most the coefficient functions $c_2$ and $c_0$ in~\eqref{eq:gamma3} seem to be affected. In light of that it is all the more remarkable that the minimal ansatz gets fully fixed up to the expected tree-level ambiguities.

\subsection{Including weight 4 functions with letter $x-\xb$}\label{sec:wider_ansatz_w4}
At weight 4, there are 3 new functions when including the additional letter $x-\xb$. Two of them are simply given by $\log(u)f^{(3)}(x,\xb)$ and $\log(v)f^{(3)}(x,\xb)$, while the third one, $f^{(4)}(x,\xb)$, is the unique such function which is fully crossing antisymmetric, just like $f^{(3)}$ at weight 3.

Including these 3 functions as additional basis elements, we now start with a new preamplitude $\P'^{(3)}$ built from an extended transcendental basis $\mathcal{Q}'$ containing 76 elements. This new, wider ansatz is again of the form
\begin{align}\label{eq:H3_wider_ansatz_w4}
	\mathcal{H}'^{(3)} = \frac{1}{u^2}(\d8)^2\,\P'^{(3)} + a_2\mathcal{H}^{(2)} + a_1\mathcal{H}^{(1)}.
\end{align}
We then proceed to impose the same bootstrap constraints described in Section~\ref{sec:constraints}. After imposing conditions (1) -- (3), the new weight 4 functions give rise to 15 additional free parameters. As before, in the process of imposing the further constraints (4) -- (6) the coefficients of the tree-level and one-loop correction terms get fixed to $a_1=-1$ and $a_2=5$. However, we now end up with 12 free parameters in the final result for $\mathcal{H}'^{(3)}$, compared to only 8 ambiguities when starting from the minimal ansatz.

We find that the four additional free parameters contribute to all antisymmetric weight~4 functions together with their lower weight completions.\footnote{
	However, note that none of these parameters is related to free coefficient $\mathcal{X}$ of reference~\cite{Huang:2021xws}. As explained in Section~\ref{sec:comparison_with_Huang/Yuan}, starting from a fully crossing symmetric preamplitude which matches the canonical form of the leading log as given by the OPE resummation excludes the function $\mathcal{X}$.
	}
In the OPE decomposition, they contribute with finite spin support to the $\log^2(u)$-part for twists $\tau\geq6$, whereas their contributions to the $\log(u)$-part have infinite spin support starting from twist 4. As a consequence, they contribute non-trivially to the two-loop anomalous dimension, but we find their contributions are limited to the coefficient function $c_0$, preserving analyticity in spin for $\ell\geq6$.

\subsection{Generalising the one-loop term}\label{sec:wider_ansatz_P2}
As another natural extension we consider a generalisation of the one-loop correlator $\mathcal{H}^{(2)}$. The minimal ansatz is thus modified by including a wider one-loop like correction term and takes the more general form
\begin{align}\label{eq:H3_wider_ansatz_P2}
	\mathcal{H}''^{(3)} = \frac{1}{u^2}(\d8)^2\,\P^{(3)} + \frac{1}{u^2}\d8\,\P^{(2)} + a_1\mathcal{H}^{(1)},
\end{align}
where $\P^{(3)}$ is the original two-loop preamplitude from~\eqref{eq:ansatz_p3}, while the preamplitude $\P^{(2)}$ is analogous to~\eqref{eq:ansatz_p3} with different coefficients $b^{(i)}_{n,m}$ which we set to zero for all basis elements $\mathcal{Q}_i$ with transcendental weight greater than 4 or contributions to $\log^3(u)$ in any orientation, in order not to spoil the matching of the leading log by the first term.

We then impose the bootstrap constraints (1) -- (3) on $\P^{(3)}$ and $\P^{(2)}$ separately, with the modification that $\P^{(2)}$ is less constrained by crossing symmetry, obeying only 
\begin{align}
	\P^{(2)}(x,\xb)=\P^{(2)}(x',\xb'),
\end{align}
since this is the only symmetry respected by a single power of $\d8$. The other symmetry is then imposed in a second step on the full combination $\frac{1}{u^2}\d8\P^{(2)}$.

Proceeding to impose the constraints (4) -- (6), we find that the coefficient of the tree-level contribution is fixed to $a_1=4$ and we have
\begin{align}
	\mathcal{H}''^{(3)} = \frac{1}{u^2}(\d8)^2\,\P^{(3)} + \frac{1}{u^2}\d8\,\P^{(2)} + 4\mathcal{H}^{(1)},
\end{align}
with $\mathcal{H}''^{(3)}$ containing a total of 11 unfixed parameters, 8 of which contribute only to $\P^{(3)}$ and these are the expected tree-level ambiguities described in Section~\ref{sec:results}. One of the additional parameters contributes only to $\P^{(2)}$ and it is proportional to the tree-level ambiguity $u^2\dbar{4444}$, which contributes to the two-loop anomalous dimension only at spin~$0$. The other 2 remaining parameters appear in both $\P^{(2)}$ and $\P^{(3)}$: one of them, $\delta_1$, contributes non-trivially up to weight 6 (with contributions to the functions $A^{(6)}$ and $\zeta_3f^{(3)}$ at top-weight) and has infinite spin support in the $\log^2(u)$-part for twists $\tau\geq6$. The other parameter, $\delta_2$, is proportional to a one-loop string correction with finite spin support in the $\log^2(u)$-part. In the two-loop anomalous dimension they both contribute with infinite spin support: $\delta_1$ contributes to $c_2$ and $c_0$, while $\delta_2$ contributes only to $c_0$. Again, we find that both contributions preserve analyticity in spin for $\ell\geq6$.\\

Lastly, when combining the two types of generalisations discussed above, i.e. starting from the extended ansatz~\eqref{eq:H3_wider_ansatz_P2} while at the same time allowing for weight 4 functions with letter $x-\xb$ (in both $\P^{(3)}$ and $\P^{(2)}$), we find 9 additional undetermined coefficients compared to the minimal ansatz. These comprise the 4 parameters described in Section~\ref{sec:wider_ansatz_w4}, the 3 from the generalisation~\eqref{eq:H3_wider_ansatz_P2} as well as 2 more independent parameters: one  of them contributes up to weight 6 and has similar properties as the parameter $\delta_1$ described above, while the second one is due to the fact that the coefficient $a_1$ is no longer fixed.

\section{Outlook and open questions}\label{sec:discussion}\setcounter{equation}{0}
We conclude with mentioning some open questions and future directions:
\begin{itemize}
	\item Recall that the simple structure of the minimal ansatz \eqref{eq:H3_ansatz_minimal} for the two-loop correlator $\mathcal{H}^{(3)}$ was motivated by two facts, namely $(i)$ the non-trivial property of the leading discontinuity allowing $(\d8)^2$ to be pulled out and $(ii)$ the observation of reference \cite{Aprile:2019rep} that the one-loop correlator (originally constructed without assuming a structure involving $\d8$) can actually be written in the simple form given in \eqref{eq:H2}. However, beyond the remarkable fact that this very restrictive minimal ansatz satisfies all of the bootstrap constraints and leads to a final result with the expected tree-level ambiguities only, the origin of this structural simplicity remains unclear to us. While the existence of the operator $\d8$ and its properties in relation to the leading log are directly related to the hidden conformal symmetry, the structures we observe at loop order seem to go beyond the original understanding of this symmetry. It would be very interesting to investigate whether analogous features, such as the appearance of lower order correlators for example, are present in similar setups.\footnote{
		One natural candidate to consider would be the case of the pure AdS$_5$ background arising from gauged $\mathcal{N}=8$ supergravity.}

	\item We find it suggestive that a similar structure might persist at higher loops. As already stated in~\cite{Huang:2021xws}, the generalisation of the minimal ansatz to any loop order takes the simple form
	\begin{align}
		\mathcal{H}^{(n)}=\frac{1}{u^2}(\d8)^{n-1}\,\P^{(n)}+\sum_{i=1}^{n-1}a_i\,\mathcal{H}^{(i)}\,,
	\end{align}
	such that at any order one is only left with determining the preamplitude $\P^{(n)}$, which includes functions up to transcendental weight $w=2n$, and the coefficients $a_i$ of the lower-loop results. Crucially, the only growth of complexity in $\P^{(n)}$ at higher loop order is due to the increasing size of the transcendental basis, while the number of free parameters in their coefficient functions does not increase. Nevertheless, the number of SVHPL's grows exponentially with their weight and considering higher loops thus remains a challenging problem.\footnote{
		Furthermore, recall that the two-loop case considered in this work is exceptional due to the accidental enhancement of crossing symmetry of $(\d8)^2$, which allowed us to start from a fully crossing-symmetric preamplitude. In general, the preamplitudes are expected to be invariant only under the symmetry $x\mapsto x'$, as one can check explicitly in the one-loop case.}

	\item For the reasons mentioned above, we expect that bootstrapping higher-loop correlators will require a better understanding of the basis of transcendental functions. In Section~\ref{sec:zigzag}, we have taken a first step in that direction by identifying the family of zigzag integrals which we argue provide a basis for the leading log at any loop-order.	On the other hand, it is clear that there is more to be understood. For example, as mentioned in Section~\ref{sec:results}, we observe that 5 basis elements at weight 6 do not contribute to the final result, which calls for an explanation. A more systematic insight into which transcendental functions do contribute at general loop orders would certainly facilitate the formulation of the most restricted ansatz for the preamplitudes $\P^{(n)}$.
		
	\item Related to that matter is the occurrence of functions containing the additional letter $x-\xb$. In Section~\ref{sec:bulk_point} we have argued that such functions are necessary to match the scale-dependent logarithmic terms of the flat-space type IIB string amplitude. In particular, such logarithms are present in the super-leading counter-terms to the supergravity contributions, and we speculate that the precise form of the corresponding counter-terms at higher-loop orders will put a bound on the maximal allowed weight of functions with letter $x-\xb$.
	
	\item An interesting immediate generalisation of the results presented here would be to consider correlators of more general external charges. As shown in~\cite{Caron-Huot:2018kta}, the ten-dimensional conformal symmetry of tree-level supergravity relates the correlator of arbitrary external charges $\langle \cO_p\cO_q\cO_r\cO_s\rangle$ to the `seed-correlator' $\langle \cO_2\cO_2\cO_2\cO_2\rangle$ through the action of a differential operator $\mathcal{D}_{pqrs}$. While this works beautifully for the free-theory and tree-level supergravity correlators, at loop order only the leading log continues to have this property. Nevertheless, as demonstrated for a number of correlators with low external charges, pulling out $\mathcal{D}_{pqrs}$ at one-loop order still achieves a great simplification, even though certain tree-level correction terms become necessary~\cite{Aprile:2019rep}. It would be fascinating to see if a similar approach can be applied to two- or even higher-loop correlators.
	
	\item In this work, we have focussed on the position space representation of $\mathcal{H}^{(3)}$. A first step to compute the corresponding Mellin amplitude has been already taken in~\cite{Bissi:2020woe}, where the contribution from the leading log $\mathcal{H}^{(3)}\vert_{\log^3(u)}$ to the two-loop Mellin amplitude has been derived. It would be instructive to derive the full Mellin amplitude and, considering that $\d8$ acts as a complicated shift-operator in Mellin space, to investigate if the simple structures observed here translate to Mellin space.
	
	\item We made the observation that the one-loop string correction $\mathcal{H}^{(2,3)}$ can in fact be written as part of the two-loop preamplitude. In other words, one can pull out $(\d8)^2$ from $\mathcal{H}^{(2,3)}$ even though it is a one-loop correlator and one would thus expect that only a single $\d8$ can be pulled out. One might wonder if all one-loop string corrections enjoy this property and what the possible consequences are.

	\item Lastly, it would be interesting to apply the position space bootstrap approach to other holographic theories beyond supergravity on AdS$_5\times$S$^5$. While there has been recent progress in this direction by constructing one-loop corrections using mainly the Mellin space formulation, e.g. for M-theory on AdS$_7\times$S$^4$ and AdS$_4\times$S$^7$ \cite{Alday:2020tgi,Alday:2021ymb} or gluon scattering on AdS$_5\times$S$^3$ \cite{Alday:2021ajh}, we believe that the position space methods employed here will continue to be useful also in different setups,\footnote{
		For example, exploiting the existence of differential operators, such as $\d8$ in our case, is more direct in the position space formulation compared to Mellin space, where differential operators act as complicated shift operators on the Mellin amplitude.}
	in particular in other instances with hidden conformal symmetry. For example, one such case is given by string theory on AdS$_3\times$S$^3$ which enjoys a six-dimensional conformal symmetry \cite{Rastelli:2019gtj,Giusto:2020neo}. To our knowledge, loop corrections to supergravity have not been constructed for this background since much less is known about the dual CFT and its spectrum.\footnote{
		See however reference \cite{Aprile:2021mvq}, where the double-trace spectrum of tensor multiplets is considered.}
	It would be interesting to investigate whether an analogous operator to $\d8$ can be used to facilitate the construction of loop corrections in that particular case.	
\end{itemize}

\section*{Acknowledgements}
We thank Francesco Aprile and Paul Heslop for initial collaboration and useful discussions at various stages of this work. HP wants to thank Alessandro Georgoudis, {\"O}mer G{\"u}rdo{\u{g}}an and Eric Perlmutter for helpful conversations. HP acknowledges support from the ERC starting grant 679278 Emergent-BH.

\appendix
\section{Details on the basis of transcendental functions}\label{app:basis}\setcounter{equation}{0}
In this appendix we spell out the details on the transcendental functions $\mathcal{Q}_i(x,\xb)$ included in our minimal ansatz as given in equations~\eqref{eq:H3_ansatz_minimal} and~\eqref{eq:ansatz_p3}. Let us start by introducing the following derivatives of the zigzag integrals $Z^{(n)}$:
\begin{align}\label{eq:derivs_zigzag}
\begin{split}
	\Psi^{(n)}(x,\xb)&=-Z^{(n)}_1(x,\xb) + (x\leftrightarrow\xb),\\
	\widetilde{\Psi}^{(n)}(x,\xb)&=-Z^{(n)}_1(x,\xb) - (x\leftrightarrow\xb),\\
	\Upsilon^{(n)}(x,\xb)&=-Z^{(n)}_2(x,\xb) + (x\leftrightarrow\xb),
\end{split}
\end{align}
where $Z^{(n)}_m(x,\xb)$ has been defined in~\eqref{eq:zigzag_derivatives}. The above combinations define pure functions, of which $\Psi^{(2)}$, $\Psi^{(3)}$, $\widetilde{\Psi}^{(3)}$ and $\Upsilon^{(3)}$ will appear in our basis. In the following, we proceed to enumerate all elements of our basis $\mathcal{Q}$ organised by their transcendental weight.

\subsubsection*{$\bullet$ Weights 0, 1 and 2}
Up to transcendental weight 2, all symmetric SVHPL's can be written as powers of logarithms of $u$ and $v$. This includes the basis elements $1$ at weight 0, the two orientations of $\log(u)$ at weight 1, and three orientations of $\log^2(u)$ at weight 2. The first antisymmetric function is given by the one-loop box function, which coincides with the first zigzag $Z^{(1)}$. Altogether, the 7 basis elements up to weight 2 read\footnote{
	Note that, for reasons of brevity, here and below we will suppress the $\xb$-argument and it is understood that all functions are really two-variable functions of $(x,\xb)$. Similarly, when denoting their transformation properties under crossing, the suppressed variable $\xb$ is transformed in the same way as its counterpart $x$, e.g. in equation~\eqref{eq:Z1_crossing}, $Z^{(1)}(1-x)$ is short for $Z^{(1)}(1-x,1-\xb)$.}
\begin{align}
	\Big\{Z^{(1)}(x),~\log^2(u),~\log^2(u/v),~\log^2(v),~\log(u),~\log(v),~1\Big\}.
\end{align}
Recall that $Z^{(1)}$ is fully crossing antisymmetric:
\begin{align}\label{eq:Z1_crossing}
	Z^{(1)}(x) = -Z^{(1)}(1-x) = -Z^{(1)}\Big(\frac{1}{x}\Big).
\end{align}

\subsubsection*{$\bullet$ Weight 3}
Within the space of SVHPL's, there are just two antisymmetric weight 3 functions, given by $\log$'s times the one-loop box function. However, as argued in the main text, we also want to include functions involving the letter $x-\xb$. Such functions make their first appearance at weight 3, with $f^{(3)}$ being the only one at this weight. Hence our antisymmetric basis elements at this weight read
\begin{align}
	\Big\{f^{(3)}(x),~\log(u)Z^{(1)}(x),~\log(v)Z^{(1)}(x)\Big\}.
\end{align}
Note that $f^{(3)}$ transforms the as $Z^{(1)}$ under crossing, obeying a similar equation as~\eqref{eq:Z1_crossing}.

On the other hand, the space of symmetric weight 3 functions (including $\zeta$-values) is given by 7 independent linear combinations of SVHPL's. These can be written as 4 orientations of $\log$'s and 3 orientations of the function $\Psi^{(2)}$ (defined as a symmetric derivative of $Z^{(2)}$, see~\eqref{eq:derivs_zigzag}):
\begin{align}
	\Big\{\log^3(u),~\log^2(u)\log(v),~\log(u)\log^2(v),~\log^3(v),~\Psi^{(2)}(x),~\Psi^{(2)}(x'),~\Psi^{(2)}(1-x)\Big\}.
\end{align}
Note that $\Psi^{(2)}$ has the symmetry $\Psi^{(2)}(x) = \Psi^{(2)}\big(\frac{1}{x}\big)$, such that only 3 independent orientations exist. Furthermore, $\zeta_3$ is implicitly included as an independent basis element thanks to the identity
\begin{align}
	\Psi^{(2)}(x) + \Psi^{(2)}(x') + \Psi^{(2)}(1-x) = 12\zeta_3.
\end{align}

\subsubsection*{$\bullet$ Weight 4}
There are 6 antisymmetric SVHPL's at this weight: 3 of them can be written as $\log$'s times $Z^{(1)}$, while the other 3 are given by orientations of the two-loop ladder integral which coincides with the zigzag function $Z^{(2)}$. The 6 antisymmetric weight 4 basis elements are hence given by
\begin{align}
\begin{split}
	\Big\{&\log^2(u)Z^{(1)}(x),~\log^2(u/v)Z^{(1)}(x),~\log^2(v)Z^{(1)}(x),\\[3pt]
	&~Z^{(2)}(x),~Z^{(2)}(x'),~Z^{(2)}(1-x)\Big\},
\end{split}
\end{align}
and we should recall that $Z^{(2)}$ obeys the crossing relation $Z^{(2)}(x)=-Z^{(2)}\big(\frac{1}{x}\big)$.

The symmetric weight 4 functions include 10 SVHPL's together with $\zeta_3$ times weight 1 functions. However, there are 3 orientations of $\log^4(u)$ which we can remove from our basis. The remaining 9 elements can be written as
\begin{align}
\begin{split}
	\Big\{&\Upsilon^{(3)}\Big(\frac{1}{x'}\Big),~\Upsilon^{(3)}(x'),~\Upsilon^{(3)}\Big(\frac{1}{1-x}\Big),~\Upsilon^{(3)}\Big(\frac{1}{x}\Big),~\Upsilon^{(3)}(x),~\Upsilon^{(3)}(1-x),\\[3pt]
	&~\log(u)\log(v)\log(uv)\log(u/v),~\log(u)\log^2(v)\log(u/v), \big(Z^{(1)}\big)^2\Big\},
\end{split}
\end{align}
with the $\zeta$-values being implicitly included within the 6 orientation of $\Upsilon^{(3)}$ thanks to the relations
\begin{align}
\begin{split}
	\Upsilon^{(3)}\Big(\frac{1}{x}\Big)+\Upsilon^{(3)}\Big(\frac{1}{x'}\Big)-\Upsilon^{(3)}(x)-\Upsilon^{(3)}(1-x) &= 12\zeta_3\log(u),\\[3pt]
	\Upsilon^{(3)}\Big(\frac{1}{1-x}\Big)~+\Upsilon^{(3)}(x')-\Upsilon^{(3)}(x)-\Upsilon^{(3)}(1-x) &= 12\zeta_3\log(v).
\end{split}
\end{align}

\subsubsection*{$\bullet$ Weight 5}
At weight 5, there is a total of 12 antisymmetric SVHPL's. Demanding no $\log^4(u)$ contributions in any channel reduces that number to 9 independent functions. We can write them as 6 orientations of the antisymmetric derivative of $Z^{(3)}$, $\widetilde{\Psi}^{(3)}$, together with 3 orientations of a function we denote by $\widetilde{\Pi}^{(5)}$. There is one more independent element given by $\zeta_3Z^{(1)}$, such that the antisymmetric weight 5 basis elements read
\begin{align}
\begin{split}
	\Big\{&\widetilde{\Psi}^{(3)}\Big(\frac{1}{x'}\Big),~\widetilde{\Psi}^{(3)}(x'),~\widetilde{\Psi}^{(3)}\Big(\frac{1}{1-x}\Big),~\widetilde{\Psi}^{(3)}\Big(\frac{1}{x}\Big),~\widetilde{\Psi}^{(3)}(x),~\widetilde{\Psi}^{(3)}(1-x),\\[3pt]
	&~\widetilde{\Pi}^{(5)}(x),~\widetilde{\Pi}^{(5)}(1-x),~\widetilde{\Pi}^{(5)}\Big(\frac{1}{x}\Big),~\zeta_3Z^{(1)}\Big\},
\end{split}
\end{align}
where $\widetilde{\Pi}^{(5)}$ is defined in terms of the following linear combination of SVHPL's
\begin{align}
\begin{split}
	\widetilde{\Pi}^{(5)}&=\L_{3,2}+\L_{2,1,2}-\L_{3,1,0}+\L_{1,2,0,0}-\L_{1,2,1,0}+\L_{2,1,0,0},
\end{split}
\end{align}
and obeys the crossing symmetry $\widetilde{\Pi}^{(5)}(x) = \widetilde{\Pi}^{(5)}(x')$.

The space of symmetric weight 5 SVHPL's is 20-dimensional. However, demanding the absence of functions with contributions to $\log^4(u)$ removes 6 degrees of freedom (these are precisely the six logarithms $\log^m(u)\log^n(v)$ at weight 5). We thus end up with 14 functions, to which we need to add $\zeta_5$ and $\zeta_3$ times symmetric weight 2 functions. Altogether, this gives 18 basis elements which we write as
\begin{align}
\begin{split}
	\Big\{&\Psi^{(3)}\Big(\frac{1}{x'}\Big),~\Psi^{(3)}(x'),~\Psi^{(3)}\Big(\frac{1}{1-x}\Big),~\Psi^{(3)}\Big(\frac{1}{x}\Big),~\Psi^{(3)}(x),~\Psi^{(3)}(1-x),\\[3pt]
	&~\Pi^{(5)}(x),~\Pi^{(5)}(1-x),~\Pi^{(5)}\Big(\frac{1}{x'}\Big),~\Pi^{(5)}(x'),~\Pi^{(5)}\Big(\frac{1}{1-x}\Big),~\Pi^{(5)}\Big(\frac{1}{x}\Big),~\Omega^{(5)},\\[3pt]
	&~\log(u)\left(Z^{(1)}(x)\right)^2,~\log(v)\left(Z^{(1)}(x)\right)^2,~\zeta_3\log^2(u),~\zeta_3\log^2(u/v),~\zeta_3\log^2(v)\Big\},
\end{split}
\end{align}
where we recognise the 6 orientations of the symmetric derivative of $Z^{(3)}$ in the first line. The definitions of $\Pi^{(5)}$ and $\Omega^{(5)}$ in terms of SVHPL's read
\begin{align}
\begin{split}
	\Pi^{(5)}&=\L_{1,2,2}-\L_{2,1,2}-\L_{2,2,0}+\L_{2,2,1}+\L_{3,0,0}+\L_{3,1,0}\\
	&~~~+\L_{1,1,2,0}+\L_{1,1,2,1}-\L_{1,2,1,0}+\L_{2,1,0,0}+2\zeta_3(4\L_{0,0}+9\L_{1,0}),
\end{split}
\end{align}
and
\begin{align}
\begin{split}
	\Omega^{(5)}&=2\L_{1,4}+\L_{3,2}+\L_{1,1,3}+2\L_{1,2,2}+\L_{1,3,1}-2\L_{2,2,0}-\L_{2,2,1}-\L_{3,1,0}\\
	&~~~-\L_{1,1,2,0}+\L_{1,2,0,0}-\L_{2,1,0,0}-2\L_{2,1,1,0}-6\zeta_3\L_{1,1},
\end{split}
\end{align}
with $\Omega^{(5)}$ being fully crossing symmetric, i.e. $\Omega^{(5)}(x)=\Omega^{(5)}(1-x)=\Omega^{(5)}\big(\frac{1}{x}\big)$.

Finally, note that $\zeta_5$ is included in the above basis as a consequence of the identity
\begin{align}
	\Pi^{(5)}(x)+\Pi^{(5)}(1-x)+\Pi^{(5)}\Big(\frac{1}{x'}\Big)+\Pi^{(5)}(x')+\Pi^{(5)}\Big(\frac{1}{1-x}\Big)+\Pi^{(5)}\Big(\frac{1}{x}\Big) = 36\zeta_5.
\end{align}

\subsubsection*{$\bullet$ Weight 6}
At top weight, we expect only antisymmetric functions to contribute, with the leading log provided by the zigzag $Z^{(3)}$. We then also include all antisymmetric weight 6 functions with no $\log^3(u)$ contribution in any orientation (given by $A^{(6)}$ and $B^{(6)}$ defined below) as well as $\zeta_3$-values times lower-weight functions (given by $\zeta_3f^{(3)}$ and $\zeta_3\log(u)Z^{(1)}$). This gives a total of 13 independent functions at weight 6, which we parametrise by
\begin{align}
\begin{split}
	\Big\{&Z^{(3)}\Big(\frac{1}{x'}\Big),~Z^{(3)}(x'),~Z^{(3)}\Big(\frac{1}{1-x}\Big),~Z^{(3)}\Big(\frac{1}{x}\Big),~Z^{(3)}(x),~Z^{(3)}(1-x),~\\[3pt]
	&~A^{(6)}(x),~B^{(6)}(x),~B^{(6)}(1-x),~B^{(6)}(x'),~\zeta_3f^{(3)},~\zeta_3\log(u)Z^{(1)},~\zeta_3\log(v)Z^{(1)}\Big\}.
\end{split}
\end{align}
The functions $A^{(6)}$ and $B^{(6)}$ are defined in terms of SVHPL's as
\begin{align}
\begin{split}
	A^{(6)}&=\L_{1,2,3}-\L_{1,3,2}-\L_{2,1,3}+\L_{2,2,2}-\L_{1,2,2,0}+\L_{1,3,1,0}+\L_{2,1,2,0}-\L_{2,2,1,0}\\
	&~~~+2\zeta_3(\L_{3}+3\L_{2,0}+2\L_{2,1}-\L_{1,0,0}+\L_{1,1,0})-15\zeta_5\L_{1},\\
\end{split}
\end{align}
and
\begin{align}
\begin{split}
	B^{(6)}&=\L_{3,3}+\L_{1,2,3}-\L_{1,3,2}+\L_{3,1,2}-\L_{1,3,0,0}-\L_{1,2,1,0,0}\\
	&~~~ +2\zeta_3(3\L_{1,2}-5\L_{2,1}+3\L_{1,0,0}+5\L_{1,1,0})-9\zeta_5\L_{1}.
\end{split}
\end{align}
Note that $A^{(6)}$ is fully crossing antisymmetric, while $B^{(6)}$ obeys one crossing symmetry:
\begin{align}
	A^{(6)}(x)=-A^{(6)}(1-x)=-A^{(6)}\Big(\frac{1}{x}\Big),\qquad B^{(6)}(x)=-B^{(6)}\Big(\frac{1}{x}\Big).
\end{align}

\section{The $\log^2(u)$ prediction at twist 4}\label{app:log^2u_twist4}\setcounter{equation}{0}
While the full $\log^2(u)$ contribution to the two-loop correlator $\mathcal{H}^{(3)}$ is not known due to mixing of exchanged operators (double-trace and potentially triple-trace contributions), restricting to the lowest twist allows one to sidestep all these problems as there is only one unique operator for each spin. The twist 4 contribution to the $\log^2(u)$ part of $\mathcal{H}^{(3)}$ can therefore be determined from lower-order CFT data. Explicitly, keeping only terms of order $\x^2$, it is given by
\begin{align}\label{eq:log2u_twist4_sum2}
	x^2g(\xb)\equiv\sum_{\ell\text{~even}}\left(\frac{1}{2}\,A_{2,\ell}^{(1)}\,\big(\gamma_{2,\ell}^{(1)}\big)^2+A_{2,\ell}^{(0)}\,\gamma_{2,\ell}^{(1)}\,\gamma_{2,\ell}^{(2)}+A_{2,\ell}^{(0)}\big(\gamma_{2,\ell}^{(1)}\big)^3 \partial_{\Delta}\right)G_{2,\ell}(x,\xb)\,, 
\end{align}
with $A^{(0)}$ and $\gamma^{(1)}$ given in equations~\eqref{eq:gamma1} and~\eqref{eq:A0}, respectively. The twist 4 next-order CFT data, i.e. the double-trace OPE coefficients $A_{2,\ell}^{(1)}$ and the one-loop anomalous dimensions $\gamma_{2,\ell}^{(2)}$, are given by~\cite{Aprile:2017bgs}
\begin{align}\label{eq:A1}
	A^{(1)}_{2,\ell} &= \left(\frac{16}{3}-128 \left(H_{\ell+3}-H_{2\ell+7}\right)-\frac{64 (2 l+9)}{2 l+7}\right)\frac{((\ell+3)!)^2 }{(2\ell+6)!}\,,\\
	\gamma^{(2)}_{2,\ell} &= \frac{1344 (\ell-7) (\ell+14)}{(\ell-1) (\ell+1)^2 (\ell+6)^2 (\ell+8)}-\frac{2304 (2 \ell+7)}{(\ell+1)^3 (\ell+6)^3} -\frac{18\alpha}{7}\,\delta _{\ell,0} \,, \label{eq:gamma2}
\end{align}
where here $H_n$ stands for the $n$-th harmonic number and $\alpha$ parametrises the contribution of the one-loop ambiguity $u^2\dbar{4444}$, which contributes only to spin $\ell=0$. Note that the correction to the OPE coefficients (for any twist $t$) can be simply obtained from the derivative relation $\langle A^{(1)}_{t,\ell}\rangle=\partial_{t}\langle A^{(0)}_{t,\ell}\gamma^{(1)}_{t,\ell}\rangle$~\cite{Heemskerk:2009pn}, where the angle-brackets $\langle\cdot\rangle$ denote averaging over the double-trace degeneracies $i=1,\ldots,t-1$.

We were able to perform the infinite sum~\eqref{eq:log2u_twist4_sum2} over even spins $\ell$ in terms of harmonic polylogarithms $H_{a}(\xb)$, resulting in
\begin{align}\label{eq:g(xb)}
\begin{split}
	g(\xb) &=  \frac{384}{5\xb^6}\Big(4\xb^2(467\xb^2-2250\xb+2250)+\xb(1145\xb^3-11266\xb^2+27000\xb-18000)H_{1}(\xb)\\
	&\qquad\qquad-2880\xb^3(\xb-2)H_{2}(\xb)-40 (\xb-1)(69\xb^3-71\xb^2-450\xb+450)H_{1,1}(\xb)\\
	&\qquad\qquad +864\xb^3(\xb-2)H_{3}(\xb)-8\xb^2(43\xb^2-450\xb+522)H_{1,2}(\xb)\\
	&\qquad\qquad +8\xb^2(151\xb^2-810\xb+774)H_{2,1}(\xb)\\
	&\qquad\qquad +96\xb^2(\xb^2-6\xb+6)(3H_{1,3}(\xb)-H_{2,2}(\xb)-2H_{3,1}(\xb))
	\Big)\\
	&\quad-\alpha\cdot\frac{384}{\xb^5}\Big(\xb(11\xb^2-60\xb+60)+3(\xb^3-12\xb^2+30\xb-20)H_{1}(\xb)\Big),
\end{split}
\end{align}
where the term in the last line is due to the spin 0 non-analyticity of the one-loop anomalous dimension $\gamma^{(2)}_{2,\ell}$ and comes with a single conformal block with $\ell=0$.


\end{document}